\documentclass[fleqn,usenatbib,useAMS]{mnras}
\usepackage[utf8]{inputenc}
\usepackage{graphicx}
\usepackage[T1]{fontenc}
\usepackage{hyperref}
\usepackage{amsmath}
\usepackage{bm}
\usepackage{amssymb}	
\usepackage[dvipsnames]{xcolor}
\usepackage{enumitem}

\newcommand*\darkU{darkU}

\usepackage{lineno}

\newcommand{\revision}[1]{\textcolor{black}{#1}}
\newcommand{\mnrasrevision}[1]{\textcolor{black}{#1}}

\newcommand{\newrevision}[1]{\textcolor{black}{#1}}

\usepackage{newtxtext,newtxmath}

\title[Detecting strongly-lensed type Ia supernovae with LSST]{Detecting strongly-lensed type Ia supernovae with LSST}

\author[N. Arendse et al.]{Nikki Arendse,$^{1}$ Suhail Dhawan,$^{2}$ Ana Sagués Carracedo,$^{1}$ Hiranya V. Peiris,$^{2,1}$ \newauthor
Ariel Goobar,$^{1}$  Radek Wojtak,$^{3}$ Catarina Alves,$^{4}$ Rahul Biswas,$^{1}$ Simon Huber,$^{5,6}$\newauthor Simon Birrer$^{7}$ and The LSST Dark Energy Science Collaboration \vspace{0.2cm} \\
$^{1}$ Oskar Klein Centre, Department of Physics, Stockholm University, SE-106 91 Stockholm, Sweden \\
$^{2}$ Institute of Astronomy and Kavli Institute for Cosmology, University of Cambridge, Madingley Road, Cambridge CB3 0HA, UK \\
$^{3}$ DARK, Niels Bohr Institute, University of Copenhagen, Jagtvej 128, 2200 Copenhagen, Denmark \\
$^{4}$ Department of Physics \& Astronomy, University College London, Gower Street, London WC1E 6BT, UK \\
$^{5}$ Max-Planck-Institut f\"ur Astrophysik, Karl-Schwarzschild Str. 1, 85741 Garching, Germany. \\
$^{6}$ Technische Universit\"at M\"unchen, Physik-Department, James-Franck-Straße 1, 85748 Garching, Germany \\
$^{7}$ Department of Physics and Astronomy, Stony Brook University, Stony Brook, NY 11794, USA
}

\date{Accepted 2024 May 9. Received 2024 April 7; in original form 2023 December 11.}

\pubyear{2023}

\begin{document}
\label{firstpage}
\pagerange{\pageref{firstpage}--\pageref{lastpage}}
\maketitle

\newcommand{\Cov}[1]{\operatorname{Cov}\left(#1\right)}

\begin{abstract}
Strongly-lensed supernovae are rare and valuable probes of cosmology and astrophysics. Upcoming wide-field time-domain surveys, such as the Vera C. Rubin Observatory's Legacy Survey of Space and Time (LSST), are expected to discover an order-of-magnitude more lensed supernovae than have previously been observed.
In this work, we investigate the cosmological prospects of lensed type Ia supernovae (SNIa) in LSST by quantifying the expected annual number of detections, the impact of stellar microlensing, follow-up feasibility, and how to best separate lensed and unlensed SNIa.
We simulate SNIa lensed by galaxies, using the current LSST baseline v3.0 cadence, and find an expected number of 44 lensed SNIa detections per year. Microlensing effects by stars in the lensing galaxy are predicted to lower the lensed SNIa detections by $\sim 8 \%$. The lensed events can be separated from the unlensed ones by jointly considering their colours and peak magnitudes. 
We define a `gold sample' of $\sim 10$ lensed SNIa per year with time delay~$> 10$ days, \revision{$> 5$ detections before light-curve peak},  and sufficiently bright ($m_i < 22.5$ mag) for follow-up observations.
In \revision{{three}} years of LSST operations, such a sample is expected to yield a $1.5\%$ measurement of the Hubble constant.
\end{abstract}




\section{Introduction}
\label{intro}

When a supernova (SN) is positioned behind a massive galaxy or cluster, it can be gravitationally lensed to form multiple images. Such an event is a rare phenomenon that can give valuable insights into high-redshift SN physics, substructures in massive galaxies, and the cosmic expansion rate. An absolute distance measurement between the observer, lens and SN can be made using the arrival time delays between the appearance of the multiple images, \revision{combined with a model for the gravitational potential of the lens galaxy and line-of-sight structures \citep{refsdal1964hubble}. This distance measure can be converted into the Hubble constant ($H_0$) -- the present-day expansion rate of the Universe.}
The exact value of the Hubble constant is an unresolved question, with different techniques yielding different results. Measurements from the cosmic microwave background (CMB) radiation \citep{planck2018cosmo} are in $5\sigma$ tension with local observations from Cepheids and type Ia supernovae (SNIa) \citep{riess2021sh0es} and from gravitationally-lensed quasars \citep{wong2020holicow}. It is worth noting that several other local measurements agree with the CMB results, such as the Tip of the Red Giant Branch (TRGB), as measured by the Carnegie-Chicago Hubble Project \citep{freedman2021perspective}, and the analysis of seven lensed quasars with less restrictive mass model priors by the TDCOSMO collaboration \citep{Birrer:2020tdcosmoIV}. 

\medskip
Strongly-lensed SNe are promising probes for obtaining a measurement of the Hubble constant which is independent of the distance ladder and early Universe physics \citep{treu2016cosmography, Suyu2023}. 
\mnrasrevision{Lensed SNe are complementary to lensed quasars and have several advantages for time-domain cosmography. 
(1) Since they are found at lower redshifts, lensed SNe have generally shorter time delays than lensed quasars, and hence, require shorter observing campaigns. 
(2) While quasar light curves are stochastic, the light curves of SNe are more predictable and there are accurate light curve templates for several subtypes, like SNIa. This makes it easier to constrain their time delays with sparse observations.
(3) When the SN fades away, follow-up observations of the lensed host galaxy light and stellar kinematics of the lensing galaxy can help to constrain the lens mass model and minimise the mass-sheet degeneracy \citep{Falco1985, Gorenstein1988, kolatt1998magnification, Saha2000, oguri2003gravitational, schneider2013MSD, birrer2021standardizable} -- a transformation of the lens potential and source plane coordinates that leaves the lensing observables unchanged. Obtaining detailed observations of the lens and host galaxy for lensed quasars is more challenging, as they are always on and often outshine the galaxies.
(4) If the lensed SN is a type Ia, their standardizable-candle nature makes them easier to identify when gravitationally magnified. Knowledge of their intrinsic brightness also helps to minimise the mass-sheet degeneracy, although recent studies show that microlensing effects from stars in the lensing galaxy complicate this by adding stochastic uncertainty to the SNIa luminosities \citep{dobler2006microlensing, Yahalomi2017, foxley2018standardization, Weisenbach2021, Weisenbach_2024}.}

\mnrasrevision{While the potential of lensed SNe for time-delay cosmography has been known for decades, these events have proved extremely difficult to detect.
}
\mnrasrevision{To date, only eight multiply-imaged lensed SNe have been discovered.}
\mnrasrevision{Six of them were found in galaxy clusters: `SN Refsdal' \citep{kelly2015multiple}}, `SN Requiem' \citep{rodney2021gravitationally}, `AT2022riv' \citep{Kelly2022riv}, \revision{`C22’} \citep{Chen2022_glSNAbell}, `SN H0pe' \citep{Frye2024}, \mnrasrevision{and SN Encore \citep{Pierel_2023_Encore}}. Furthermore, two SNIa have been discovered lensed by a single elliptical galaxy: `iPTF16geu' \citep{goobar2017iptf16geu} and `SN Zwicky' \citep{Goobar2023_SNZwicky}.
\mnrasrevision{SN Refsdal and SN H0pe yielded Hubble constant measurements with uncertainties of $\sim 6\%$ \citep{kelly2015multiple} and $\sim 9\%$ \citep{Pascale2024_H0pe_H0}, respectively, where the precision was primarily limited by the cluster mass models. While SN Encore is expected to enable an $H_0$ measurement with $\sim 10\%$ precision,} the other lensed SNe had either insufficient data or too-short time delays to usefully constrain the Hubble constant.

\medskip
The study of lensed SNe is currently at a turning point, as we will go from a handful of present discoveries to an order-of-magnitude increase with the next generation of telescopes such as the Vera C. Rubin Observatory \citep{ivezic2019lsst} and the Nancy Grace Roman Space Telescope \citep[e.g.,][]{pierel2020projected}. In particular, the Legacy Survey of Space and Time (LSST) to be conducted at the Rubin Observatory is predicted to discover several hundreds of lensed SNe per year according to studies based on limiting magnitude cuts \citep{wojtak2019magnified,goldstein2017glSNe, oguri2010gravitationally}.

\medskip
In this work, we take into account the full LSST baseline v3.0 observing strategy to quantify the expected number of lensed SNIa detections per year and investigate the properties of the predicted sample. \mnrasrevision{We focus on lensed SNIa because they are especially promising for cosmology by virtue of their standardizable-candle nature and predictable light curves.}
We examine the colours and apparent magnitudes of the simulated lensed SNIa sample and study how to best separate them from unlensed SNIa. Finally, we measure time delays from simulated \newrevision{light curves with only LSST data} and show how to construct a `gold sample’ which is promising for follow-up observations. Our results include the effects of microlensing due to stars in the lens galaxy. Throughout this work, we assume a standard flat $\Lambda$CDM model with \revision{$H_0 = 67.8 \ \textrm{km} \, \textrm{s}^{-1} \textrm{Mpc}^{-1}$ and $\Omega_{\textrm{m}} = 0.308$ \citep{13planck2015}.}

\medskip
We outline our lensed SNIa simulation procedure in Sec.~\ref{sect2:sims}, the LSST observing strategy in Sec.~\ref{sect3:rubin}, and the different aspects of detecting lensed SNIa in Sec.~\ref{sect4:detecting}. Our results are presented in Sec.~\ref{sect:results} and our discussion and conclusions in Sec.~\ref{sect:conclusions}.


\section{Simulating lensed SNIa}
\label{sect2:sims}

\noindent In this work, we develop the publicly available Python code called \textit{lensed Supernova Simulator Tool} (\texttt{lensedSST}\footnote{\href{https://github.com/Nikki1510/lensed_supernova_simulator_tool}{https://github.com/Nikki1510/lensed\_supernova\_simulator\_tool}}), which simulates a sample of lensed SNIa light curves with the LSST baseline v3.0 observing strategy to perform our analysis.
Additionally, we simulate a sample of unlensed SNIa as a ``background" population, to compare their colours to lensed SNIa. Ancillary catalogue information such as the Einstein radius, \newrevision{SN image positions,} time delays between images and magnifications of the lens systems are also saved and used in our work. 
In this section, we describe our assumptions in terms of the lens galaxy mass model, SNIa light curves, and microlensing simulations.

\begin{table}
\begin{center}
\renewcommand{\arraystretch}{1.2}
\begin{tabular}{ l l}
\hline \hline
{\bfseries Parameter} & {\bfseries Distribution} \\
\hline
Hubble constant & \revision{$H_0 = 67.8 \ \textrm{km} \, \textrm{s}^{-1} \textrm{Mpc}^{-1}$} \\
Lens redshift & $z_{\textrm {lens}} \sim \mathcal{N}^{ \mathcal{S}}(3.88, 0.13, 0.36)$ \\
Lensed source redshift & $z_{\textrm {src}} \sim \mathcal{N}^{ \mathcal{S}}(3.22, 0.53, 0.55)$ \\
Unlensed source rate & $r_{v} (z) = 2.5 \cdot 10^{-5} (1+z)^{1.5} \,\textrm{Mpc}^{-3}\textrm{yr}^{-1}$ \\
Source position (doubles) & $x_\textrm{src}, y_\textrm{src} \sim \mathcal{U}(-\theta_{\textrm{E}}, \theta_{\textrm{E}})$ \\
Source position (quads) & $x_\textrm{src}, y_\textrm{src} \sim \mathcal{U}(-0.4\theta_{\textrm{E}}, 0.4\theta_{\textrm{E}})$ \\
 & \\
{\bfseries Lens galaxy} \\
\hline 
{Elliptical power-law mass profile} \\
{Lens centre $(^{\prime \prime})$} & $x_\textrm{lens}, y_\textrm{lens} \equiv (0, 0)$ \\
Einstein radius $(^{\prime \prime})$ & $\theta_{\textrm{E}} \sim \mathcal{N}^{ \mathcal{S}}(5.45, 0.14, 0.63)$ \\ 
Power-law slope & $\gamma_{\textrm {lens}} \sim \mathcal{N}(2.0, 0.2)$ \\ 
Axis ratio & $q_{\textrm {lens}} \sim \mathcal{N}(0.7, 0.15)$ \\ 
Orientation angle (rad) & $\phi_{\textrm {lens}} \sim \mathcal{U}(-\pi/2, \pi/2)$ \\ 
 & \\
{\bfseries Environment} \\
\hline 
External shear modulus & $\gamma_{\textrm {ext}} \ \sim \mathcal{U}(0, 0.05)$ \\
 & \\
{\bfseries Light curve} \\
\hline 
Stretch & $x_1 \ \sim \mathcal{N}^{ \mathcal{S}}(-8.24, 1.23, 1.67)$ \\
Colour & $c \ \sim \mathcal{N}^{ \mathcal{S}}(2.48, -0.089, 0.12)$ \\
Absolute magnitude & $M_{\mathrm{B}} \ \sim \mathcal{N}(-19.43, 0.12)$\\
Milky Way extinction & $E(B-V) \ \sim \mathcal{U}(0, 0.2)$ \\
\hline \hline
\end{tabular}
\end{center}
\caption{\label{tab:param_distributions} \textbf{Parameter distributions for lensed SNIa.} The distribution of input parameters employed in the simulation pipeline to generate lensed SNIa light curves. $\mathcal{N}(\mu, \sigma)$ indicates a normal distribution with mean $\mu$ and standard deviation $\sigma$, $\mathcal{N}^{\mathcal{S}}(a, \mu, \sigma)$ denotes a skewed normal distribution with skewness parameter $a$, while $\mathcal{U}(x, y)$ represents a uniform distribution with bounds $x$ and $y$. The skewed normal distributions for $z_{\textrm{lens}}$, $z_{\textrm{src}}$ and $\theta_{\textrm{E}}$ are 1D representations of the full joint distribution \revision{from \citet{wojtak2019magnified}} depicted in Fig.~\ref{fig:zl_zs_thetaE_distribution}.}
\end{table}

\subsection{Lens galaxy mass profile assumptions}

We employ the multi-purpose lens modelling package \texttt{Lenstronomy}\footnote{\href{https://lenstronomy.readthedocs.io/en/latest/}{https://lenstronomy.readthedocs.io/en/latest/}} \citep{birrer2018lenstronomy, Birrer2021LenstronomyII} to generate lens galaxies with a power-law elliptical mass distribution (PEMD) to describe the projected surface mass density, or convergence $\kappa$:
\begin{align}
    \kappa(x, y) = \frac{3 - \gamma_{\rm lens}}{2} \left(\frac{\theta_E}{\sqrt{q_\textrm{lens} x^2 + y^2/q_\textrm{lens}}} \right)^{\gamma_{\rm lens} - 1},
    \label{eq:pemd}
\end{align}
where $q_\textrm{lens}$ is the projected axis ratio of the lens, $\gamma_{\rm lens}$ corresponds to the logarithmic slope, and $\theta_E$ denotes the Einstein radius. The coordinates ($x, y$) are centred on the position of the lens centre, and rotated by the lens orientation angle $\phi_\textrm{lens}$, such that the $x$-axis is aligned with the major axis of the lens. We model the external shear from the line-of-sight structures with a shear modulus $\gamma_{\textrm{ext}}$ and a shear angle $\phi_{\textrm{ext}}$, which we assume to be uncorrelated from the lensing galaxy orientation. The adopted parameter distributions are given in Table~\ref{tab:param_distributions}. 

\revision{\subsection{Redshift distribution}}
\label{subsect:sims:redshift}

\begin{figure}
	\centering
		{\includegraphics[width=0.45\textwidth,clip=true]{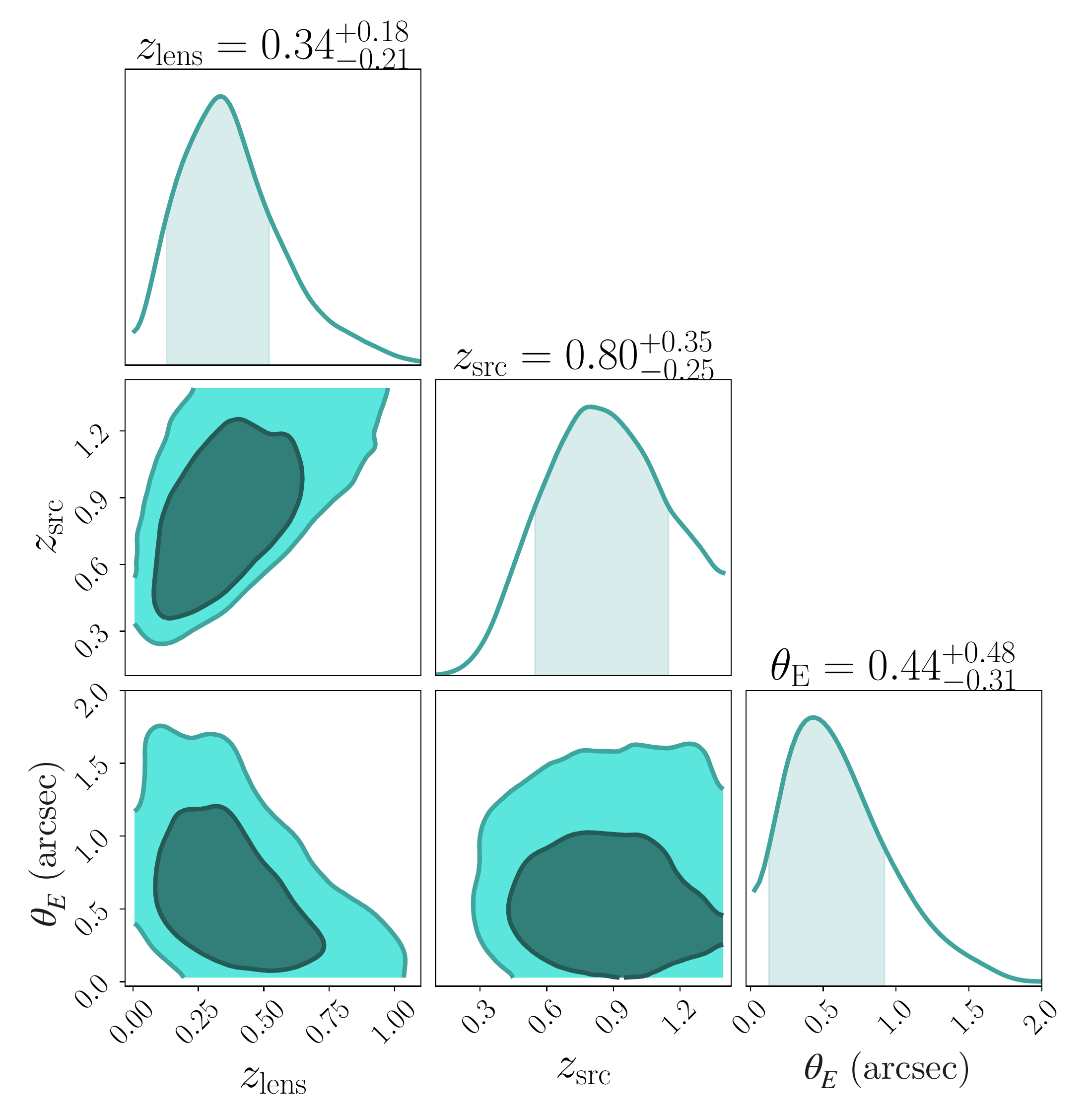}}
	\caption{The joint distribution of the lens redshift ($z_{\textrm{lens}}$), source redshift ($z_{\textrm{src}}$) and Einstein radius ($\theta_{\textrm{E}}$) used to simulate the lensed SNIa systems. The $z_{\textrm{lens}}$, $z_{\textrm{src}}$ and $\theta_{\textrm{E}}$ combinations correspond to galaxy-source configurations where strong lensing occurs \citep{wojtak2019magnified}. The sample includes all lensed SNIa at redshift $z_{\rm src}<1.5$.}
	\label{fig:zl_zs_thetaE_distribution}
\end{figure}

\noindent Most of our input parameters are uncorrelated, except for the Einstein radius $\theta_{\textrm{E}}$, lens redshift $z_{\textrm{lens}}$ and source redshift $z_{\textrm{src}}$, \newrevision{which we sample from a kernel density estimation probability model fitted to ($\theta_{\textrm{E}}$,  $z_{\textrm{lens}}$,  $z_{\textrm{lens}}$) of gravitationally lensed SNIa generated in \citet{wojtak2019magnified}.} The \newrevision{aforementioned} simulation assumes a population of lens galaxies with the velocity dispersion function derived from the Sloan Digital Sky Survey observations \citep{choi2007galaxies} and a model of the volumetric rate of SNIa fitted to measurements of the SNIa rate as a function of redshift \citep{rodney2014rate}. \newrevision{The lensed SNIa rate is determined by drawing random realizations of lens galaxies and supernovae in a light cone and counting the events where strong lensing occurs. More details can be found in \citet{wojtak2019magnified}.} 
We impose an additional upper limit on the source redshift of $z_{\textrm{src}} < 1.5$ to ensure that the SNe are not redshifted out of the filters. The resulting combinations of $z_{\textrm{lens}}$, $z_{\textrm{src}}$ and $\theta_{\textrm{E}}$ values are depicted in Fig.~\ref{fig:zl_zs_thetaE_distribution} \revision{and the projected 1D distributions can be found in Table~\ref{tab:param_distributions}}.

\begin{figure}
	\centering
		{\includegraphics[width=\hsize,clip=true]{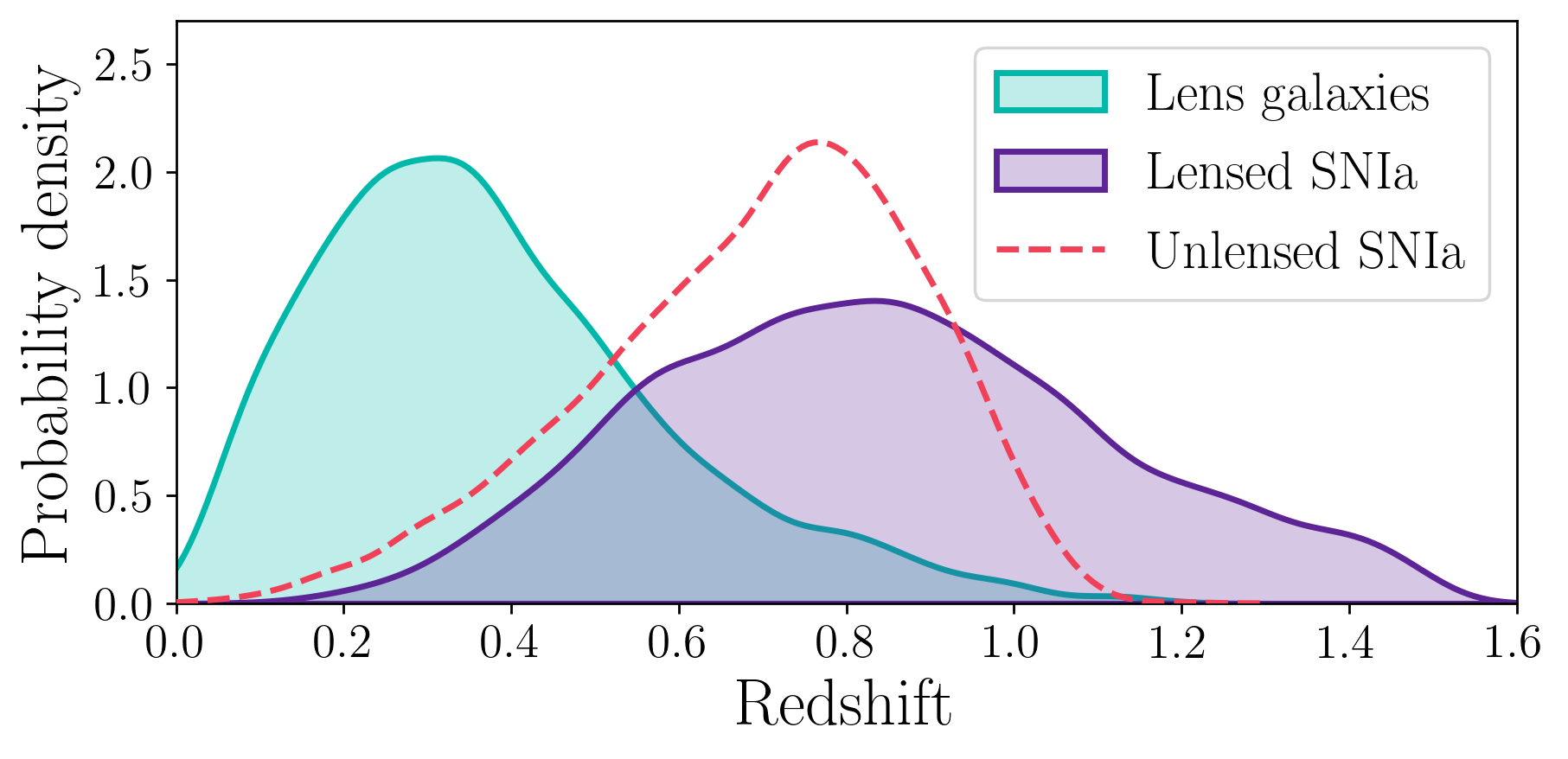}}
	\caption{Normalised redshift distributions of simulated lens galaxies, lensed SNIa and unlensed SNIa that will be detectable with the Rubin Observatory. \revision{The Figure shows the observed populations, after the detection criteria described in Sec.~\ref{subsect:sims:redshift} for unlensed SNIa and in Sec.~\ref{subsect:predicted_glSNe_rates} for lensed SNIa have been applied.} }
	\label{fig:z-distr}
\end{figure}

\medskip
For the background population of unlensed SNIa, we assume a volumetric redshift rate of \citep{Dilday2008}
\begin{align}
r_{v} (z) = 2.5 \cdot 10^{-5} (1+z)^{1.5} \ \textrm{Mpc}^{-3} \; \textrm{yr}^{-1}.
\end{align}
\revision{Our sample of detected unlensed SNIa consists of objects with $i$-band magnitude brighter than 24.0.
For lensed SNIa we adopt stricter detection criteria, which are described in Sec.~\ref{subsect:predicted_glSNe_rates}.} The resulting redshift distributions of detected lensed and unlensed SNIa and lens galaxies in LSST are displayed in Fig.~\ref{fig:z-distr}. The Rubin Observatory will be able to discover unlensed SNe at redshifts up to \revision{$z \sim 1$}, resulting in a significant overlap in redshift space between lensed and unlensed SNIa. \\

\subsection{SNIa light curves}

We model the SNIa as point sources, using synthetic light curves in the observer frame for their variability. The light curves are simulated using \texttt{SNCosmo}\footnote{\href{https://sncosmo.readthedocs.io/en/stable/}{https://sncosmo.readthedocs.io/en/stable/}}\citep{barbary2016sncosmo} and its in-built parametric light curve model \revision{\texttt{SALT3} \citep{guy2007salt2, Kenworthy2021}}, which takes as input an amplitude parameter $x_0$, stretch parameter $x_1$, and a colour parameter $c$. We sample the $x_1$ and $c$ parameters from asymmetric Gaussian distributions that have been derived by \citet{scolnic2016measuring} for the Supernova Legacy Survey \citep{guy2010SNLS}, the Sloan Digital Sky Survey \citep{sako2018SDSS}, Pan-STARRS1 \citep{rest2014cosmological}, and several low-redshift surveys \citep{Hicken2012, Stritzinger2011}. We compute \revision{the distance modulus $\mu$ of each SNIa, based on its $x_1$ and $c$ parameters:
\begin{align}
&\mu = \newrevision{m} + \alpha  x_1 - \beta  c - \mathcal{N}(M_0, 0.12),
\label{eq:cosmo_correction}
\end{align}}
\noindent\revision{Here, $M_0$ is the expected absolute magnitude of a SNIa with $x_1 = c = 0$ and \newrevision{$m$ is the SN flux normalisation in magnitude units.} We assume $M_0 = -19.43$ in the $B$-band, corresponding to a universe with Hubble constant $H_0 = 67.8 \ \textrm{km} \, \textrm{s}^{-1} \textrm{Mpc}^{-1}$.} 
\revision{$\alpha$ and $\beta$ are the linear stretch and colour correction coefficients, as first found in \citet{Phillips1993} and \citet{tripp1998colorcorrection} respectively, which specify the correlation of absolute magnitude with the stretch and colour parameters. We assume $\alpha = 0.14$ and $\beta = 3.1$ \citep{scolnic2016measuring}}. The resulting absolute magnitude values for each SNIa are used as input for \textsc{SNCosmo} to generate the corresponding unlensed light curves.

\medskip
\revision{The final, lensed light curves are computed in the following way. After drawing $z_{\textrm{lens}}$, $z_{\textrm{src}}$ and $\theta_{\textrm{E}}$ from the joint distribution from \citet{wojtak2019magnified} (Fig.~\ref{fig:zl_zs_thetaE_distribution}), we sample random positions of the source and the remaining lens parameters from the distributions given in Table~\ref{tab:param_distributions} until we find a system that is detectable by LSST. The criteria for what we consider as a `detection' are described in Sec.~\ref{subsect:predicted_glSNe_rates} for lensed SNIa. Using \texttt{Lenstronomy}, we obtain the image positions, time delays and magnifications for each lensed SNIa. The time delays and magnifications are applied to the unlensed SNIa light curve to obtain the final, lensed light curves.} 

\begin{figure*}
	\centering
		{\includegraphics[width=\textwidth,clip=true]{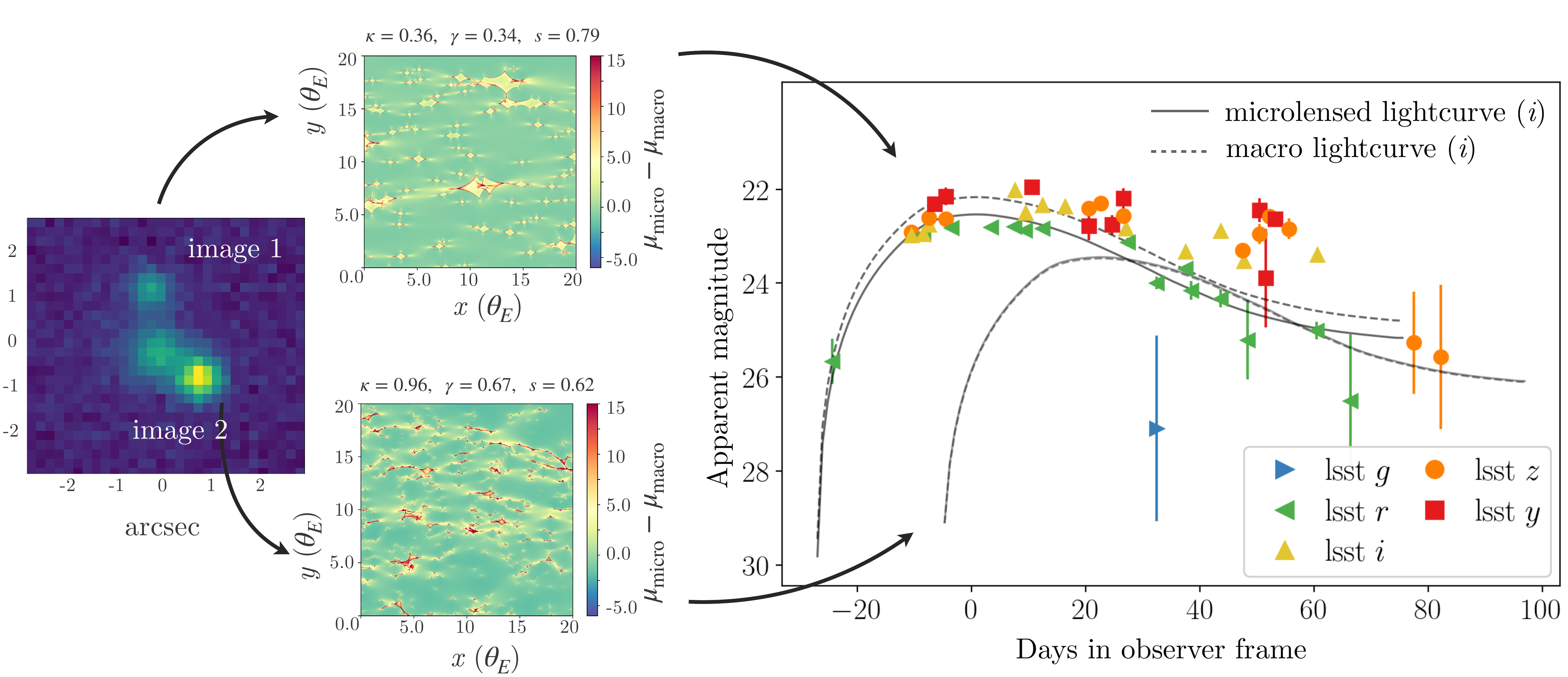}}
	\caption{Simulated observations of a doubly imaged SNIa in LSST. For each lensed SN image, the figure shows the microlensing magnifications maps and the corresponding $i$-band light curves in the observer frame (dashed curves without microlensing and solid curves with microlensing). The coloured markers correspond to LSST observations in the active region of the WFD survey with the baseline v3.0 cadence. The properties of this simulated lensed SN system are $z_{\rm lens} = 0.1, z_{\rm src} = 0.52, \theta_{\rm E} = 1.26'', \Delta t = 22$ days. }
	\label{fig:lightcurves}
\end{figure*}

\subsection{Microlensing}

Stars (and dark matter substructures) in the lens galaxy can give rise to additional gravitational lensing effects on top of the lens' macro magnification. Such \textit{microlensing} effects from stars are typically able to magnify or demagnify the lensed SN images by approximately one magnitude. Since the resulting microlensing magnifications are not symmetric -- some systems will be highly magnified while the majority will be slightly demagnified -- their effects can change the number of lensed SNe that will pass our detection thresholds. We included microlensing in our simulations and investigated the resulting impact on the annual lensed SNIa detections.

\medskip
To calculate microlensed light curves we follow the approach as described by \cite{Huber2019LSSTcadence}, where synthetic observables from theoretical SNIa models
calculated via \texttt{ARTIS} \citep{Kromer:2009ce} are combined with microlensing magnification maps. The maps are generated following \citet{Chan2021} and with software from {\tt GERLUMPH} \citep{Vernardos:2014lna,Vernardos:2014yva,Vernardos:2015wta}.
As in \cite{huber2021timedelay} we create maps with a Salpeter initial mass
function with a mean mass of the microlenses of \revision{$\langle M \rangle = 0.35$} $\mathrm{M}_\odot$, a resolution of 20000
$\times$ 20000 pixels and a total size of \revision{20 $R_{\rm E}$ $\times$ 20 $R_{\rm E}$. Here, $R_{\rm E}$ corresponds to the physical Einstein radius of the microlenses at the source redshift and can be calculated via
\begin{equation}
    R_{\rm E} = \sqrt{\frac{4G \langle M \rangle}{c^2} \frac{D_{\rm s} D_{\rm ls}}{D_{\rm l}}},
\end{equation}
where $D_{\textrm{l}}$, $D_{\textrm{s}}$ and $D_{\textrm{ls}}$ are the angular diameter distances between the observer and the lens, the observer and the source, and the lens and the source, respectively.}
Further, we list in Table~\ref{table:microlensing} the convergence $\kappa$, the shear $\gamma$ and the smooth matter
fraction $s$ ($s=1-\kappa_*/\kappa$, where $\kappa_*$ is the convergence of
the stellar component) for all magnification maps considered in this work. \revision{The specific realisations of $\kappa$, $\gamma$ and $s$ were chosen because they correspond to the most commonly occurring combinations amongst the simulated lensed SNIa.
We normalise the microlensing magnification maps to have the same mean as the theoretical magnification predicted from the map's $\kappa$ and $\gamma$ values:
\begin{equation}
    \mu = \frac{1}{(1-\kappa^2) - \gamma^2}.
\end{equation}
}
For each map we have 40000 microlensed spectra coming from 10000 random positions in the map and four theoretical SN models, the same as used by \cite{Suyu:2020opl,Huber:2020dxc} and \cite{huber2021timedelay}. For all the maps listed in Table~\ref{table:microlensing} we assumed a source redshift of 0.77 and a lens redshift of 0.32, which corresponds to the median values of the OM10 catalog \citep{Oguri:2010} and defines the total size of the map \revision{$R_{\rm E}$}. For our lensed SNe we are interested in \revision{$z_{\mathrm{src}}$} between 0.0 and 1.4\revision{. To reduce the computational effort we grid the $z_{\mathrm{src}}$ space} in steps of 0.05. Given that the calculation of 10000 microlensed spectra for a single magnification map with a certain \revision{$R_{\rm E}$} is on the order of a week, we approximate the microlensing contributions, as we now describe. For any source redshift of interest $z_{\mathrm{src}}$ we use the microlensed spectra calculated for the source redshift of 0.77. We then rescale the spectra such that they correspond to $z_{\mathrm{src}}$ in terms of absolute flux, wavelength and time after explosion. From the corrected spectra we can then calculate the exact light curves for $z_{\mathrm{src}}$ following \cite{Huber2019LSSTcadence}, with the approximation that the total size of the microlensing map is the same as for the source redshift of 0.77. Using the same total size can slightly overestimate or underestimate the impact of microlensing, but \cite{Huber:2020dxc} tested different \revision{$R_{\rm E}$} values, where no significant dependence between the strength of microlensing and \revision{$R_{\rm E}$} was found.

\medskip
For each simulated lensed SN, we compute the convergence, shear and smooth matter fraction at the position of the SN images and draw a random microlensing realisation from the magnification map with the closest $\kappa$, $\gamma$ and $s$ values. The local convergence and shear are calculated from the lens galaxy's mass model and the smooth matter fraction is obtained by approximating the stellar convergence $\kappa_{*}$ at the image positions, for which we assume a spherical de Vaucouleurs profile \citep{dobler2006microlensing}: 
\begin{equation}
     \kappa_{*} (r) = A e^{-k(r/R_{\textrm{eff}} )^{1/4}},
\end{equation}
where $k = 7.67$, $r$ is the radius of the SN image position to the lens centre, $R_{\textrm{eff}}$ is the effective radius of the lens and $A$ is a \newrevision{normalisation} constant that is calibrated for each lens system such that the maximum $s$ value is 1. The effective radius $R_{\textrm{eff}}$ of the lens is determined through a scaling relation between the radius \revision{in $\textrm{kpc} \ \textrm{h}^{-1}$} and velocity dispersion $\sigma$ \revision{in $\textrm{km} \ \textrm{s}^{-1}$} of elliptical galaxies \citep{Hyde2009curvature}:
\begin{align}
    &\revision{\log_{10}(R_{\rm eff})} = 2.46 - 2.79 \cdot \log_{10}(\sigma) + 0.84 \cdot \log_{10}(\sigma)^2 \\
    &\textrm{with } \ \sigma^2 = \frac{c^2 \, \theta_{\rm E} \, D_{\rm s}}{4  \pi \, D_{\rm ls}},
\end{align}
The above relations also assume spherical symmetry of the lens galaxy, but are only used to determine which $\kappa$, $\gamma$ and $s$ values are the best approximations for the image positions. 

\medskip
Finally, the microlensing contributions are obtained by drawing a random position in the chosen magnification map for each lensed SN image. Since the SN explosion models comprise different sizes at different wavelengths, the chromatic microlensing contributions are computed for each LSST filter and added to the simulated lensed SN light curves, as illustrated in Fig.~\ref{fig:lightcurves}.

\begin{table}
\centering
\begin{tabular}{c|c|c}
\hline \hline
Convergence ($\kappa$) & Shear ($\gamma$) & Smooth matter fraction ($s$)     \\
\hline
0.362                 & 0.342                 & 0.443 \\
0.655                 & 0.669                 & 0.443 \\
0.655                 & 0.952                 & 0.443 \\
0.956                 & 0.669                 & 0.443 \\
0.956                 & 0.952                 & 0.443 \\
0.362                 & 0.342                 & 0.616 \\
0.655                 & 0.669                 & 0.616 \\
0.655                 & 0.952                 & 0.616 \\
0.956                 & 0.669                 & 0.616 \\
0.956                 & 0.952                 & 0.616 \\
0.362                 & 0.342                 & 0.790 \\
0.655                 & 0.669                 & 0.790 \\
0.655                 & 0.952                 & 0.790 \\
0.956                 & 0.669                 & 0.790 \\
0.956                 & 0.952                 & 0.790 \\
0.362                 & 0.280                 & 0.910 \\
\hline \hline
\end{tabular}
\caption{Combinations of the convergence ($\kappa$), shear ($\gamma$), and smooth matter fraction ($s$) used to simulate the microlensing contributions to the lensed SN light curves.}
\label{table:microlensing}
\end{table}

\section{The Vera C. Rubin Observatory} 
\label{sect3:rubin}

\noindent The Vera C. Rubin Observatory is a survey facility currently under construction on Cerro Pachón in Chile. It will host the Legacy Survey of Space and Time (LSST), a wide-field astronomical survey scheduled to start operations around 2025. The survey will take multi-colour $ugrizy$ images and cover $\sim$20,000 square degrees of the sky in a ten-year period. Due to its depth and sky coverage, LSST is the most promising transient survey for observing gravitationally lensed SNe, with initial predicted numbers of several hundred discoveries a year \citep{goldstein2017glSNe, goldstein2019rates, wojtak2019magnified, oguri2010gravitationally}. 

\subsection{LSST observing strategy}

LSST will operate several survey modes. The main programme, comprising \revision{$\sim$ 90\%} of observing time, will be the Wide-Fast-Deep (WFD) survey, consisting of an area of 18,000 square degrees. The other major survey programmes include the Galactic plane, polar regions and the Deep Drilling Fields (DDFs); the latter will be observed with deeper coverage and higher cadence. \newrevision{Following the latest recommendations from the Survey Cadence Optimization Committee\footnote{\href{https://pstn-055.lsst.io/}{https://pstn-055.lsst.io/}},} the WFD survey is expected to proceed using a \textit{rolling cadence}, in which certain areas of the WFD footprint will be assigned more frequent visits, with the focus of increased visits ``rolling'' over time. This improves the light curve sampling of the objects discovered in those high-cadence areas, the \textit{active} regions. The drawback is that since the sky coverage is not homogeneous in any given period, there is a greater chance of missing rare events if they occur in an under-sampled area, the so-called \textit{background} region. For unlensed SNe, it has been shown by \citet{Alves2022} that the active region yields a $25\%$ improvement in type classification performance relative to the background region. \newrevision{One of the goals of this work} is to determine the impact of the rolling cadence on lensed SNIa. \citet{Huber2019LSSTcadence} \newrevision{investigated the effects of several LSST observing strategies on lensed SNIa discoveries}, but this previous work considered earlier versions of the observing strategy \newrevision{that differed significantly} from the current implementation of the rolling cadence. 

\begin{figure*}
	\centering
	{\includegraphics[width=0.95\textwidth,clip=true]               {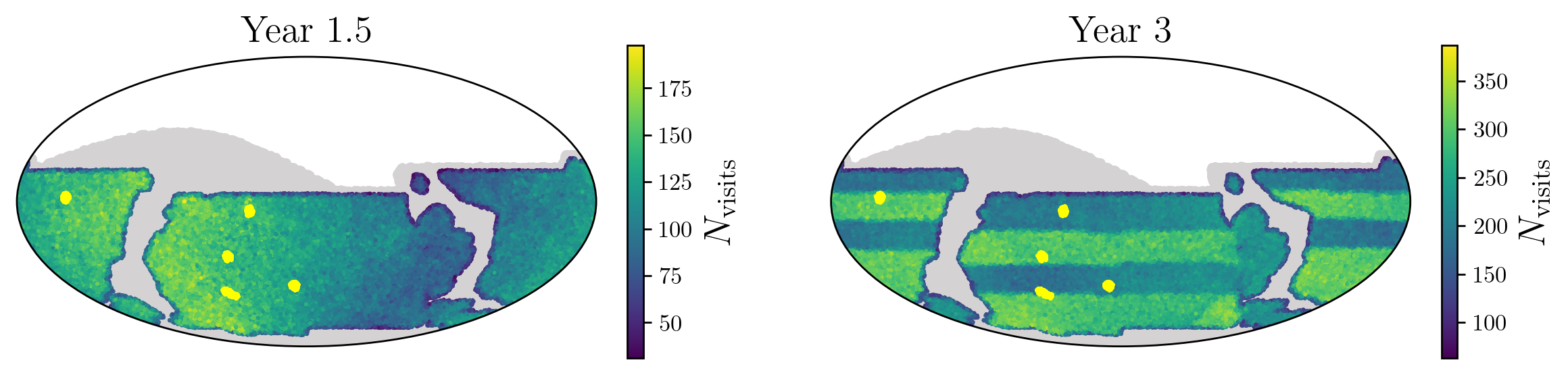}}
	\caption{LSST survey footprint in equatorial coordinates showing the number of visits ($N_{\rm visits}$) conducted per sky location. The light gray area corresponds to the full LSST footprint, including the Galactic plane and polar regions. The coloured region represents the WFD area ($\sim 18,000$ sq. deg.) and the light yellow patches are the DDFs. In the first 1.5 years of the survey (left panel) the sky coverage will be homogeneous, after which the rolling cadence will start (right panel). Rolling results in background regions with lower $N_{\rm visits}$ and active regions with higher $N_{\rm visits}$. }
	\label{fig:sky_dist}
\end{figure*}

\medskip
\newrevision{We implement the baseline v3.0 observing strategy, which} adopts a half-sky rolling cadence with a $\sim 0.9$ rolling weight (corresponding to the background regions receiving only 10\% of the standard number of visits, and the active regions the rest). \newrevision{While the baseline v3.3 cadence was recently released, we do not expect it to significantly alter our results. The biggest change with respect to v3.0 is an updated mirror coating which results in decreased $u$-band sensitivity and increased sensitivity in the $grizy$ bands. Since lensed SNIa are red and our analysis does not consider $u$-band detections, the change from v3.0 to v3.3 is minimal, and only expected to result in slightly more lensed SNIa detections.} 
A sky map with the observations corresponding to the baseline v3.0 strategy is depicted in Fig.~\ref{fig:sky_dist}. The rolling cadence begins roughly 1.5 years after the start of the survey, to first allow for a complete season of uniform observations. This gives us an ideal opportunity to compare the effects of the rolling cadence on the discovery of lensed SNIa. For the annual lensed SN detections computed in this work, we only consider observations in the WFD and the DDFs, since those do not suffer from severe dust extinction found e.g. in the Galactic plane regions. An example of a lensed SNIa with baseline v3.0 WFD cadence observations is depicted in Fig.~\ref{fig:lightcurves}.

\subsection{Simulating LSST observations}
\label{subsect:simulating_LSST}

In order to simulate observations of lensed SNe at the catalogue level with sufficient information to define useful metrics, we need to find the set of times at which SNe at particular locations are observed, along with the observational metadata required to estimate the uncertainty with which the SN flux will be measured. 
This information can be accessed through the Rubin Operations Simulator (\texttt{OpSim}), which simulates the field selection and image acquisition process of LSST over the 10-year duration of the planned survey \citep{2016SPIE.9910E..13D,2014SPIE.9150E..15D,2019AJ....157..151N}. 
Detailed information about each simulated pointing of the telescope is stored in an output data product in the form of a `sqlite` database, where each pointing forms a row in the database. Using \texttt{OpSimSummary}~\citep{2020ApJS..247...60B}, we find all the observations (and associated metadata) that include the position of a given SN within the field of view of the telescope. 

\medskip
In order to separate the Galactic plane region and the WFD and DDF surveys, we use a threshold based on the number of visits a sky location has received after 10 years of LSST observations. This serves as a proxy for \texttt{OpSim}'s distinction between the Galactic plane and WFD regions. We assume that regions with $N_{\rm visits,10yr} < 400$ belong to the Galactic plane and polar regions, $N_{\rm visits,10yr} > 1000$ are the DDFs, and all remaining sky locations are assigned to the WFD. Within the WFD, we distinguish between observations taken during the non-rolling phase (year $0$--$1.5$; MJD $< 60768$) and the first rolling period (year $1.5$--$3$; $60768 <$ MJD $< 61325$).

\medskip
From the \texttt{OpSim} database, we obtain the observing times, filters, \revision{mean 5-sigma depth ($m_5$)}, and point-spread functions (psf). \revision{We compute the $1 \sigma$ noise on the SN flux (which contains contributions from the sky brightness, the airmass, the atmosphere, and the psf) from the 5-sigma depth in the following way:} 

\begin{equation}
    \revision{\sigma_{f_{\rm SN}}} = \frac{10^{(\textrm{ZP} - \revision{m_5})/2.5}}{5},
\end{equation}
where ZP corresponds to the \revision{instrument zero-point for a given band: $28.38, 28.16, 27.85, 27.46, 26.68$, respectively for the $g, r, i, z, y$ bands \footnote{\href{https://smtn-002.lsst.io/\#change-record}{https://smtn-002.lsst.io/\#change-record}}.}
The SN flux for each image is perturbed by drawing a new flux value from a normal distribution with mean of the model flux and width equal to the sky noise. Then, the new flux values $f_{\rm SN}$ are combined with the \revision{SN flux noise} to calculate the error on the observed magnitude:
\begin{equation}
    \left| \frac{-2.5 \cdot \revision{\sigma_{f_{\rm SN}}}}{f_{\rm SN} \cdot \textrm{ln}(10)} \right|.
\end{equation}

\section{Detecting lensed SNIa} 
\label{sect4:detecting}

\noindent In this section, we describe our methods for calculating the number of lensed SNIa detections and the properties of the detected sample. We examine the simulated lensed SNIa based on their colours, magnitudes, time-delay measurements, and prospects for follow-up.

\subsection{Annual lensed SNIa detections}
\label{subsect:predicted_glSNe_rates}

Studies that predict the number of lensed SN discoveries generally take into account two distinct detection methods. The first is the \textit{image multiplicity method}, which looks for multiple resolved images of the lensed SN \citep{oguri2010gravitationally}, and the second is the \textit{magnification method}, which looks for objects that appear significantly brighter than a typical SN at the redshift of the lens galaxy (which acts as the apparent host galaxy) \citep{goldstein2017glSNe}. For the latter method, the lensed SN images do not need to be resolved.
We build our estimates of the lensed SN discoveries upon the results from \cite{wojtak2019magnified}. They combine the image multiplicity and the magnification method and predict that LSST will discover around 89 lensed SNIa per year, assuming a 0.2~mag buffer above $5\sigma$ average limiting magnitudes. \revision{For comparison, the predicted rate of unlensed SNe~Ia, after quality cuts for cosmological utility, is 104000 for the ten-year sample. \citep{lsst_srd}}

\medskip
The number of lensed SN detections from \cite{wojtak2019magnified} considers whether a lensed SN passes the image multiplicity and magnification cuts based on its full light curve information. There is no observation cadence information included in the predictions, which would alter the results if a lensed SNe will occur between observing seasons or when sufficient observations are missing around the peak. Here, we update the annual lensed SNIa detections taking into account the cadence from the baseline v3.0 observing strategy. For each simulated lensed SN, we draw a random observation sequence and assess whether the object still passes the detection cuts for the magnification and image multiplicity method.

\medskip
The criteria for being ``detected'' with both methods are the following: \\

\noindent \textit{Image multiplicity method}
\begin{itemize}
    \item The maximum image separation $\theta_{\rm max}$ is larger than $0.5''$ and smaller than $4''$;
    \item For doubles: the flux ratio between the images \newrevision{(measured at peak)} is between 0.1 and 10;
    \item At least three or two images are detected (signal-to-noise ratio $> 5$) for quads (systems with four images) and doubles (systems with two images), respectively.
\end{itemize}

\noindent \textit{Magnification method}
\begin{itemize}
    \item The apparent magnitude of the unresolved lensed SN images should be brighter than a typical SNIa at the \textit{lens} redshift at peak:
    \begin{equation}
        m_{\rm X} (t) < \langle M_{\rm X} \rangle (t_{\rm peak}) + \mu (z_{\rm lens}) +
K_{\rm XX}(z_{\rm lens}, t_{\rm peak}) + \Delta m,
    \end{equation}
    with $m_{\rm X} (t)$ the apparent magnitude of the transient in band X at time $t$, $\langle M_{\rm X} \rangle (t_{\rm peak})$ the absolute magnitude of a standard SNIa in band X at peak, $\mu$ the distance modulus, $K_{\rm XX}$ the K-correction, and $\Delta m$ the magnitude gap (adopted here to be $-0.7$ \revision{for consistency with \cite{wojtak2019magnified} and \cite{goldstein2017glSNe}});
    \item The combined flux of the unresolved data points should be above the detection threshold (signal-to-noise ratio $> 5$).
\end{itemize}

\begin{figure*}
    \centering
    \includegraphics[width=.85\textwidth]{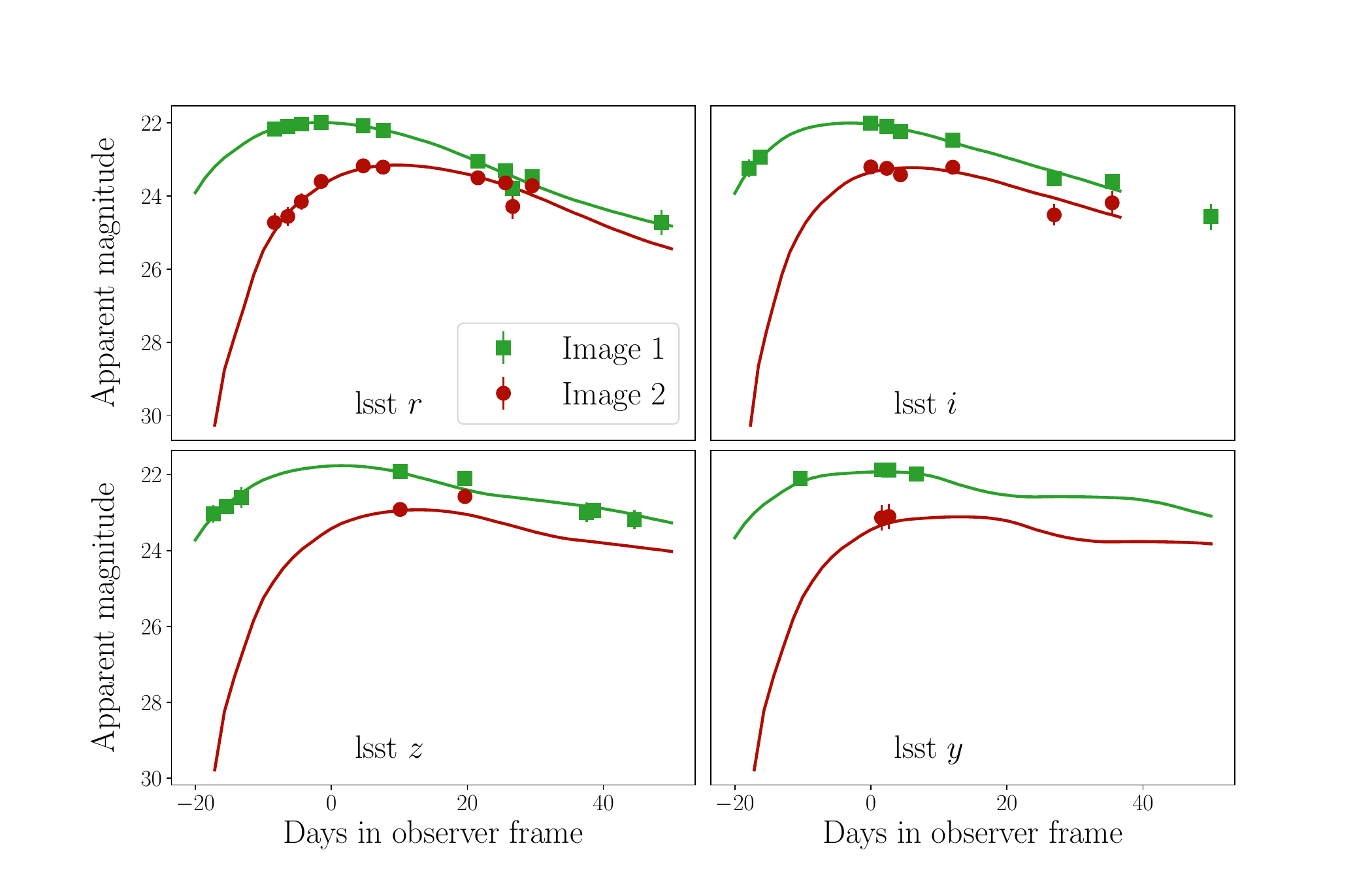}
        \caption{Example of a lensed SNIa with a robust time-delay inference from LSST data only. The markers correspond to LSST observations taken in the $r,i,z,y$ bands and the solid curves show the \revision{\texttt{SALT3}} fits used to infer the time delay. This object is at $z_{\rm src} = 0.464$, lensed by a deflector galaxy at $z_{\rm lens} = 0.142$, an input time delay of 11.14 days and a recovered time-delay estimate of 12.23 days. The Einstein radius of the system is $\theta_{\textrm E} = 0.63$ so the images are treated as resolved.}
    \label{fig:glSN_dt_saltfit}
\end{figure*}

\subsection{Colours and magnitudes of lensed SNIa}
\label{subsect:methods:colour-mag}

SNe affected by strong gravitational lensing are expected to look different from unlensed SNe in several ways. Fig.~\ref{fig:z-distr} shows that in general, lensed SNe will be found at higher redshifts than unlensed ones and hence, they will be observed as redder and more slowly evolving. Additionally, for lensed and unlensed SNIa at the same redshifts, the lensed SNIa will appear brighter because of the gravitational lensing magnification.

\medskip
In this analysis, we investigate which observables are best suited to separate the populations of lensed and unlensed SNIa in LSST data.  We aim to investigate optimal selection criteria based on the brightness and colours of lensed SN candidates. We measure light curve properties in the observer frame from the simulated sample of lensed and unlensed SNIa as observed with LSST. The resulting observables are apparent magnitudes for each LSST filter ($g$,$r$,$i$,$z$,$y$) and all colour combinations ($g-r$, $g-i$, $g-z$, $g-y$, $r-i$, $r-z$, $r-y$, $i-z$, $i-y$ and $z-y$), at different epochs \revision{at the light curve peak, which is determined in the following way.} A polynomial fit is performed on every light curve of the sample to find the peak time from the filters with the best detection cadence, which mostly corresponds to the $r$ or $i$ bands. We use the same polynomial fits to obtain the expected apparent magnitudes at the given epochs to compute the colours. Error bars from the detections are considered in the fits and propagated into uncertainties on the measured magnitudes and colours. 

\medskip
The redshift distributions for unlensed and lensed SNIa observed with LSST are largely overlapping, as shown in Fig.~\ref{fig:z-distr}. Due to this, using the apparent magnitude and colours alone will not serve as good metrics to separate lensed from unlensed SNIa, as is also illustrated in Fig.~\ref{appendixfig:mag_all} and \ref{appendixfig:colour_all}. Therefore, we chose to investigate cuts based on all combinations of colours versus apparent magnitudes. \newrevision{In \citet{quimby2014detection}, such a colour-magnitude cut in the $r$ and $i$-band is shown to be very promising for distinguishing lensed SNIa.}
We aim to devise linear cuts in colour-magnitude space for all LSST bands that exclude most of the unlensed events while preserving the lensed ones. Our method consists of obtaining the 90\% contour for the unlensed SNe in each colour-magnitude space and fitting a linear function to this contour to extract a simple linear cut.

\subsection{Time-delay measurements}
For a fraction of the lensed SN discoveries, LSST will be able to resolve the individual images. Here, we compute the fraction of those systems for which we will be able to infer the time delay precisely using LSST data only. Even for events that are on the cusp of being resolvable and with variable seeing, a single epoch with sufficient seeing to resolve the images will enable the extraction of the full light curves using forced photometry. To find the objects with the best time-delay measurements, we limit ourselves to systems with an angular separation of $\theta_E > 0.5$". We use the most commonly implemented model for the SED of an SNIa, \revision{\texttt{SALT3} \citep{guy2010SNLS, Betoule2014, Kenworthy2021}} \newrevision{as implemented in \texttt{SNCosmo} \citep{barbary2016sncosmo}} and fit each light curve with a common stretch and colour parameter. Since our aim is to infer what fraction of objects have an accurately estimated time delay, we do not use a simultaneous inference for extinction and magnification like for iPTF16geu \citep{Dhawan2020_16geu} or SN~Zwicky \citep{Goobar2023_SNZwicky}. We assume the SN redshift will be known from spectroscopic follow-up observations, which can be carried out after the SN has faded away. The difference in returned $t_0$ values provides the \mnrasrevision{predicted} time delay between the images, \mnrasrevision{$\Delta_{t, \textrm{pred}}$}.  We classify a system as having a ``good" time delay when \mnrasrevision{$\left| \Delta_{t, \textrm{pred}} - \Delta_{t, \textrm{true}} \right| / \Delta_{t, \textrm{true}} < 5\%$}. An example case is shown in Fig.~\ref{fig:glSN_dt_saltfit}. We only use LSST data \revision{for  $\Delta t$ measurements} and hence our results are a conservative estimate which will improve further with follow-up observations, especially if the follow-up also resolves the individual images. 

\subsection{Gold sample for follow-up observations}
\label{subsect:methods:goldsample}

In order to conduct timely follow-up observations of the lensed SNe, we should find them early on in their evolution. After the initial detection of a lensed SN candidate in LSST, spectroscopic follow-up observations are needed to verify its lensed nature. A spectrum will reveal the SN type and its redshift, which for SNIa will identify the objects that are magnified by strong gravitational lensing. From  simulated detected lensed SNe, we construct a `gold' sample that satisfies the following criteria:
\begin{itemize}
    \item $N_{\rm premax} > 5$ in at least two filters;
    \item $m_i < 22.5 $ mag;
    \item $\Delta t > 10$ days,
\end{itemize}
with $N_{\rm premax}$ the number of detections with signal-to-noise ratio > 3 before the SN peak and $m_i$ the apparent $i$-band magnitude at peak.

\medskip
We use \revision{\texttt{SALT3}}, \newrevision{as implemented in \texttt{SNCosmo} \citep{barbary2016sncosmo}} to fit the light curves of the simulated lensed SNIa and infer the time of peak. To trigger spectroscopic follow-up observations we require that the object is detected in at least two filters and has a minimum of five observations before the inferred time of maximum. This is because it allows for spectroscopic follow-up when the lensed SN is close to its brightest, while still having ample time for scheduling high-resolution follow-up. In addition, we apply the constraint that the lensed SNe should be bright enough to get a classification spectrum with a 4-m class telescope, e.g. 4MOST \citep{Swann2019} or e.g. the New Technology Telescope \citep{Snodgrass2008} with exposure times $\lesssim$~1hr, corresponding to a brightness of $m_i < 22.5$ mag. \revision{Alternatively, with shorter exposure times, the spectra can be obtained with instruments on 8m class telescopes, e.g. the Gemini Multi-Object Spectrographs \citep[GMOS;][]{Crampton2000}.}


\begin{figure}
	\centering
		{\includegraphics[width=\hsize,clip=true]{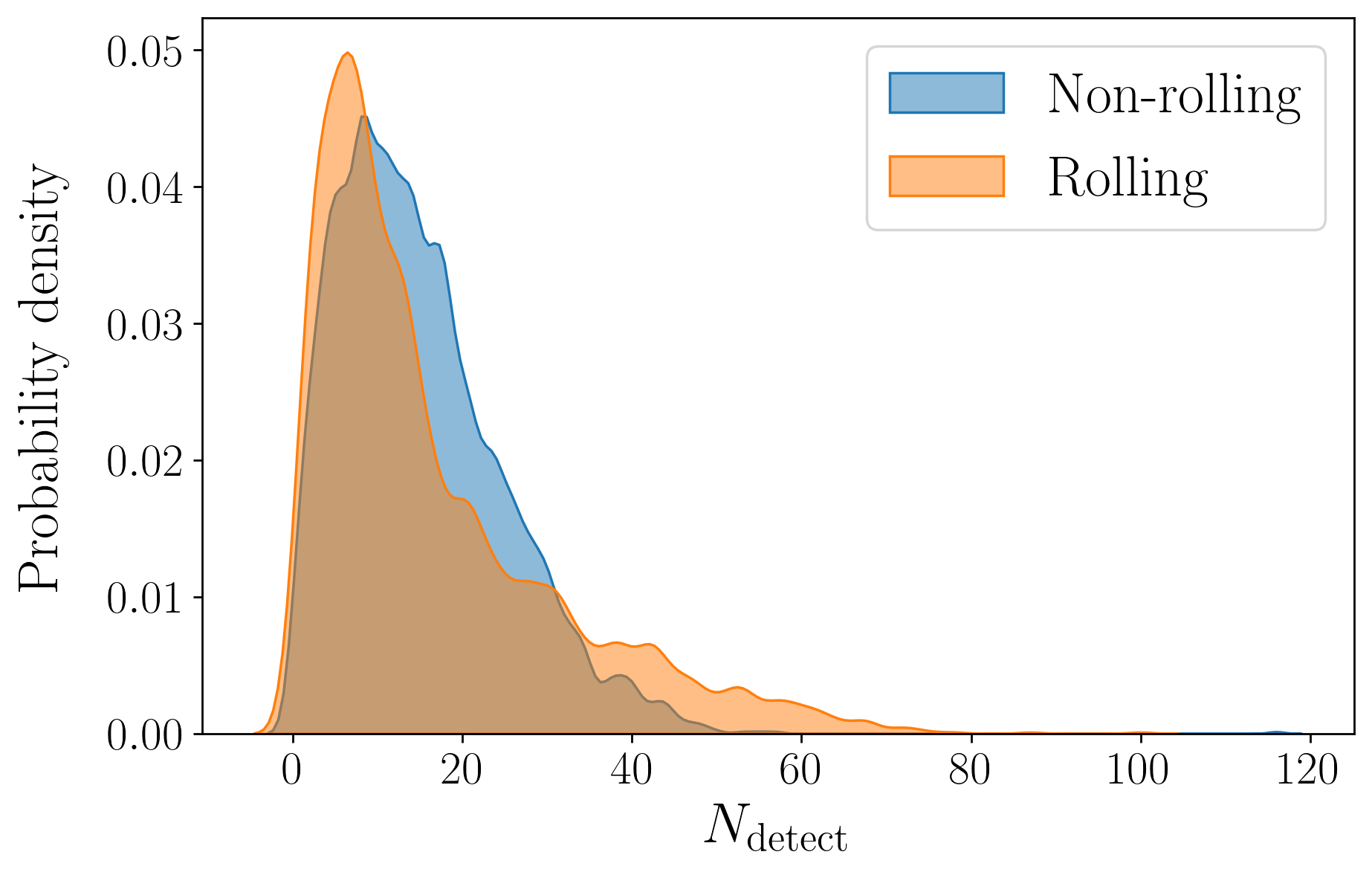}}
	\caption{Number of \newrevision{detections ($N_{\textrm{detect}}$)} per lensed SNIa for the non-rolling cadence (first 1.5 years of the survey) compared to the rolling cadence (years 1.5 - 3 of the survey). The distributions show that compared to the non-rolling cadence, the rolling cadence produces both more systems with small $N_{\textrm{detect}}$ and with large $N_{\textrm{detect}}$. \revision{The mean number of observations is 15 (17) for a non-rolling (rolling) cadence.}}
	\label{fig:Nobs}
\end{figure}

\begin{figure}
	\centering
		{\includegraphics[width=\hsize,clip=true]{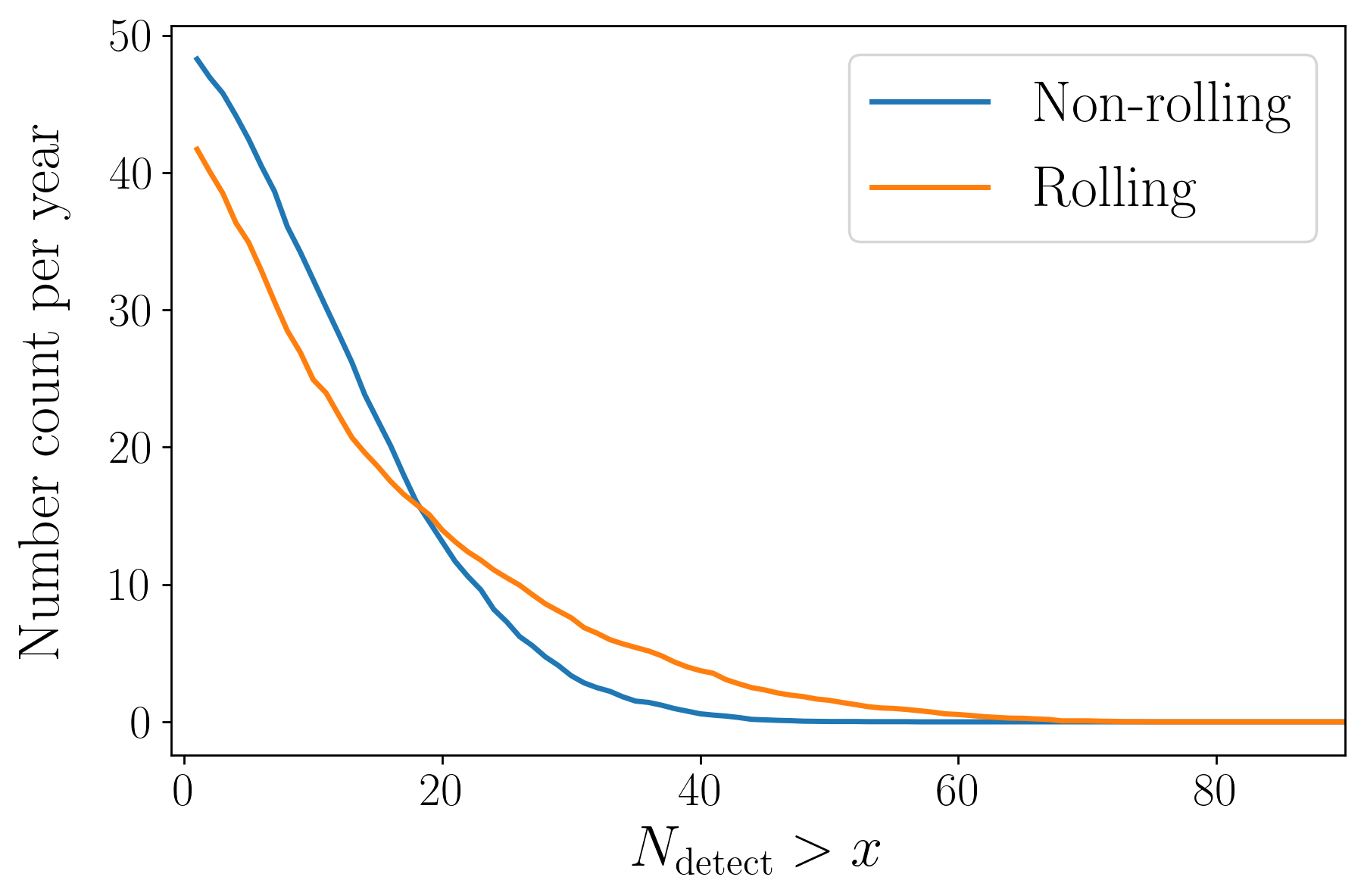}}
	\caption{Annual expected number of lensed SNIa with \newrevision{$N_{\textrm{detect}}$} above a certain threshold, for observing strategies without a rolling cadence (blue line) and with rolling cadence (orange line). The figure shows that although non-rolling cadences are expected to discover a larger overall number of lensed SNe, rolling cadences will provide more lensed SNe with a high number of detections.}
	\label{fig:N_glsne}
\end{figure}

\section{Results}
\label{sect:results}

\subsection{Annual lensed SNIa detections}

\noindent We simulate a set of $5,000$ doubly imaged lensed SNe and $5,000$ quadruply imaged SNe that pass either the image multiplicity or the magnification method as described in Sec.~\ref{subsect:predicted_glSNe_rates}\revision{, where the size of the sample is chosen such that it is large enough for our statistical analysis. \revision{These numbers are subsequently scaled with the predicted lensed SNIa rates from \citet{wojtak2019magnified}: 64 doubles and 25 quads per year.}
We then compute what fraction of the simulated lensed SNIa} remains detectable when the baseline v3.0 observing strategy is applied. We find that $46\%$ of doubles and $70\%$ of quads remain from the full simulated sample. 
We also assess the impact of the rolling cadence on the annual lensed SNIa detections, by separating the sample into objects that are detected in the first 1.5 years of the survey (MJD < 60768) and objects discovered in years 1.5 to 3 (60768 < MJD < 61325). 

\medskip
Fig.~\ref{fig:Nobs} shows the number of \newrevision{detections with signal-to-noise ratio > 5 ($N_{\rm detect}$)} per lensed SN system for the rolling and non-rolling cadence. The distribution for a non-rolling baseline v3.0 cadence peaks around 20 \newrevision{detections} per lensed SN, while the rolling cadence has a large tail towards systems with higher numbers of \newrevision{detections.} The background regions (corresponding to the dark areas in Fig.~\ref{fig:sky_dist}) acquire a lower number of detections, while the active regions (light areas in Fig.~\ref{fig:sky_dist}) receive a higher \newrevision{$N_{\rm detect}$.} As a result, the non-rolling cadence scans a larger area of the sky with a medium cadence and therefore discovers more lensed SNe, while the rolling-cadence provides more detections for the systems it discovers. This effect is illustrated in Fig.~\ref{fig:N_glsne}, which shows the expected number of lensed SNIa per year with \newrevision{$N_{\rm detect}$} above a certain threshold. The non-rolling cadence will discover more lensed SNe up to \newrevision{$N_{\rm detect} < 20$}, while the rolling cadence will find a larger sample with well-sampled light curves (\newrevision{$N_{\textrm{detect}}$ > 20}). Nevertheless, we note that the differences are relatively small. We also compute the number of lensed SNIa that fall in the Deep Drilling Fields (DDFs) in our simulation, since those objects will be observed with a much higher cadence and better depth. However, we find that only $\sim 0.2$ lensed SNIa per year are expected to be in the DDFs, which is not surprising given the small area covered relative to WFD.

\medskip
Our findings are summarised in Table~\ref{tab:results}, which contains the predicted annual number of lensed SNIa detections for a non-rolling versus a rolling cadence. \newrevision{Our results are consistent with a recent study by \citet{SainzdeMurieta2023}, which predicts the number of lensed SNIa detected with the magnification method for an approximate LSST survey strategy.}

\begin{table}
\begin{center}
\renewcommand{\arraystretch}{1.2}
\begin{tabular}{l | cc}
\hline \hline
\textbf{Doubles}    &   Non-rolling & Rolling \\
\hline
Detected without microlensing                     & 36          & 31      \\
Detected  & 32          & 27      \\
\mnrasrevision{Detected with mag. method} & \mnrasrevision{13} & \mnrasrevision{12} \\
\mnrasrevision{Detected with im. mult. method} & \mnrasrevision{24} & \mnrasrevision{19} \\
with $\Delta t > 10$ days         & 22          & 18      \\
Pass colour-mag cut         & 13         & 11    \\
Gold sample         & 8          & 6      \\
                                   &             &         \\
\textbf{Quads}                              &             &         \\
\hline
Detected without microlensing                     & 18          & 17      \\
Detected  & 17          & 16      \\
\mnrasrevision{Detected with mag. method} & \mnrasrevision{13} & \mnrasrevision{13} \\
\mnrasrevision{Detected with im. mult. method} & \mnrasrevision{13} & \mnrasrevision{11} \\
with $\Delta t > 10$ days         & 8          & 8       \\
Pass colour-mag cut         & 7        & 7      \\
Gold sample         & 5          & 4      \\
                                   &             &         \\
\textbf{Total }                             &             &         \\
\hline
Detected without microlensing                    & 54          & 48      \\
\textbf{Detected}  & \textbf{50}          & \textbf{44}      \\
\mnrasrevision{Detected with mag. method} & \mnrasrevision{26} & \mnrasrevision{25} \\
\mnrasrevision{Detected with im. mult. method} & \mnrasrevision{37} & \mnrasrevision{31} \\
with $\Delta t > 10$         & 30          & 26    \\
Pass colour-mag cut         & 22          & 19    \\
Gold sample         & 13          & 10      \\
\hline \hline
\end{tabular}
\end{center}
\caption{\label{tab:results} The annual expected number of discovered lensed SNIa (doubles, quads, and the total number) in the baseline v3.0 observing strategy, with separate predictions for the non-rolling and rolling cadence. The rows list the predicted numbers of lensed SNIa that are detected without and with microlensing (see \ref{subsect:results:microlensing}), \mnrasrevision{are detected with the magnification (mag.) method and image multiplicity (im. mult.) method,}
have time delays larger than 10 days, pass the colour-magnitude cut (described in \ref{subsect:results:colour_mag}), and are in the gold sample for follow-up (\ref{subsect:results:goldsample}). }
\end{table}

\subsection{Microlensing impact}
\label{subsect:results:microlensing}

\noindent \revision{While all results presented so far included the effects of microlensing,} we also generate each lensed SN light curve both with and without microlensing in order to clearly quantify the microlensing impact for each system.  We distinguish three scenarios, in which microlensing effects
\begin{enumerate}[noitemsep,topsep=0pt, leftmargin=0.7\parindent]
    \item do not change the detectability of the lensed SN; 
    \item make a detected lensed SN undetectable;
    \item make an undetected lensed SN detectable.
\end{enumerate}

\begin{figure}
	\centering
		{\includegraphics[width=1.05\hsize,clip=true]{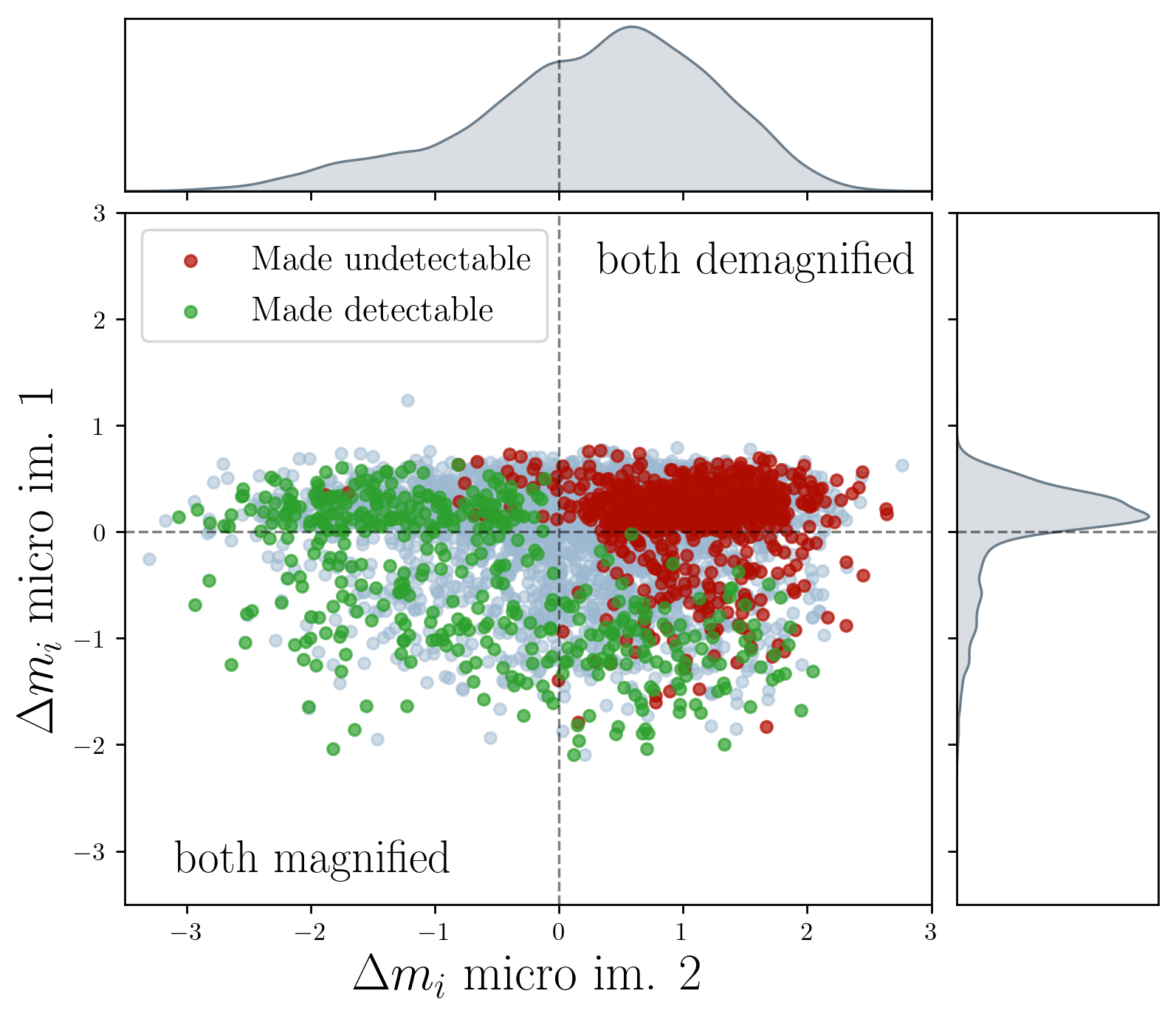}}
	\caption{Impact of microlensing on the lensed SNIa detections. The scatterplot shows the difference in apparent peak $i$-band magnitude ($\Delta m_i$) due to microlensing for image one (first occurring) and two of the 5,000 simulated doubly-imaged SNe. The gray points correspond to lensed SNe whose detection is not impacted by microlensing, the red points are the systems that have become undetectable by microlensing, and the green ones have become detectable by a microlensing magnification boost. The projected 1D distributions show the microlensing effect on the apparent magnitude per lensed SN image.}
	\label{fig:microlensing}
\end{figure}

\medskip
Fig.~\ref{fig:microlensing} investigates these three scenarios. It shows the difference in apparent peak $i$-band magnitude for the 5,000 simulated doubly-imaged SNe. The red dots are the systems that become undetectable because of microlensing, while the green dots are the ones that have become detectable due to the microlensing magnifications. The sum of these effects is that we detect a handful fewer lensed SNe; Table~\ref{tab:results} presents that we go from a total annual number of 48 (54) without microlensing to 44 (50) with microlensing for a rolling (non-rolling) cadence. \revision{For the events with longer time delays than 10 days, we predict to find 29 (31) without microlensing and 26 (30) with microlensing for a rolling (non-rolling) cadence.} This weak \revision{effect of detecting fewer objects when microlensing is included in the simulations} can be understood when looking at the projected 1D distributions of Fig.~\ref{fig:microlensing}, which shows that the majority of events will be slightly demagnified due to microlensing, while a rare few will be highly magnified. 
\mnrasrevision{Note also the asymmetry in the distributions. The first image is a minimum of the time delay and can never be demagnified -- microlensing can therefore only affect it subject to this constraint. The second image is a saddlepoint with no such restriction on its magnification and can therefore be more severely influenced by microlensing \citep{Schechter_Wambsganss_2002}.}

\begin{figure}
	\centering
		{\includegraphics[width=1.01\hsize,clip=true]{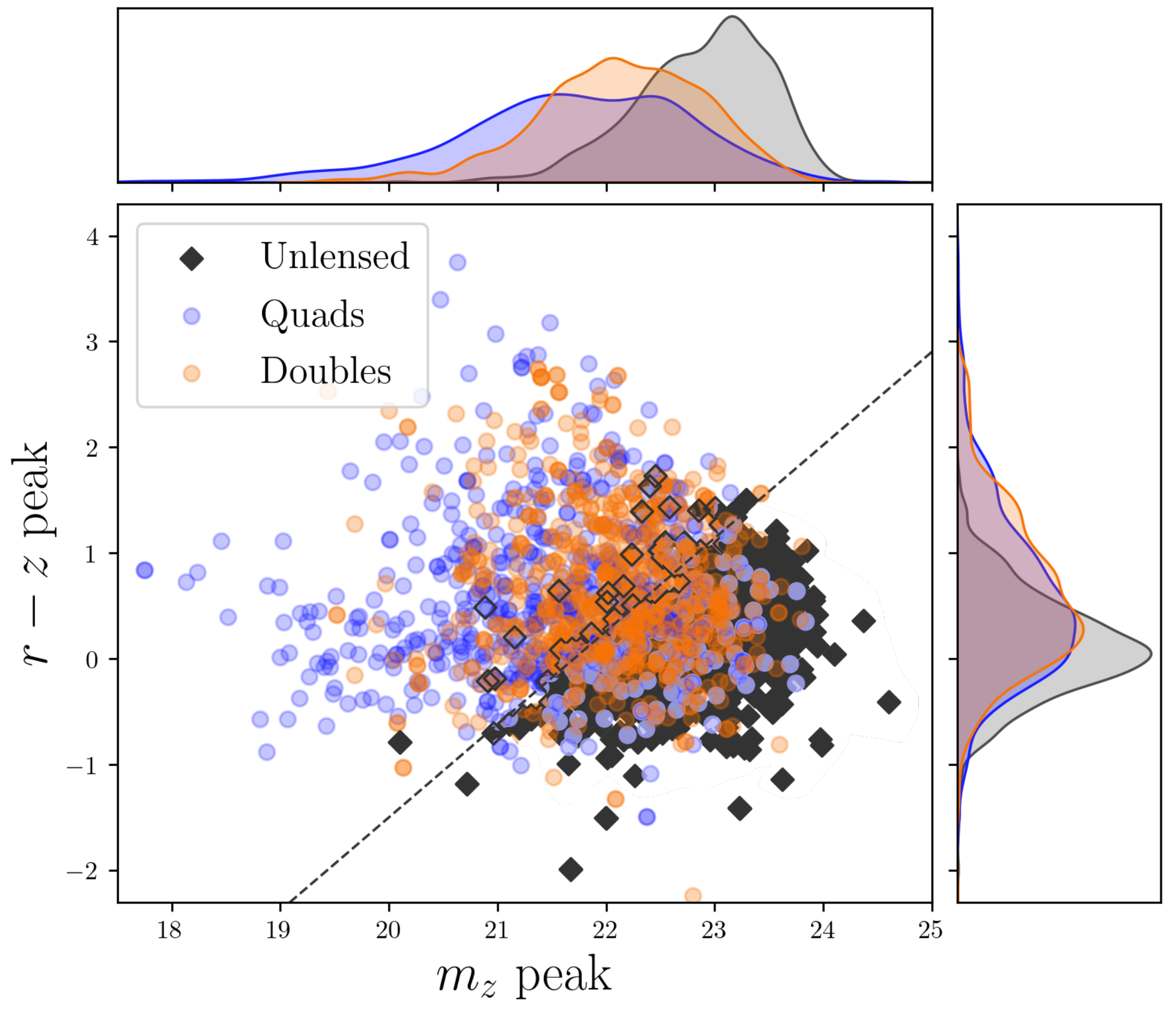}}
	\caption{Peak colours and magnitudes of the detected lensed SNIa (doubles and quads) and unlensed SNIa. The $r-z$ colours and $z$-band magnitudes are able to separate the populations because lensed SNe are brighter than unlensed ones at similar redshifts. The dashed black line shows a simple linear separation cut of $r-z > 0.88 \, m_z - 19.1$.}
	\label{fig:colour-mag}
\end{figure}

\subsection{Colour and magnitudes of lensed and unlensed SNIa}
\label{subsect:results:colour_mag}

For each of the simulated lensed and unlensed SNIa, we calculate the apparent magnitudes at peak in every band, following the procedure outlined in Sec.~\ref{subsect:methods:colour-mag}. The best separation between lensed and unlensed SNIa is achieved with the $r-z$ peak colour versus observed apparent $z$-band peak magnitude, which is shown in Fig.~\ref{fig:colour-mag}. Other colour and magnitude combinations are included in Appendix~\ref{appendix:colour-mag}. Due to their higher redshift distributions, lensed SNe are expected to appear redder than unlensed ones. However, since LSST will also detect unlensed SNe at high redshifts (see Fig.~\ref{fig:z-distr}), this difference is less pronounced in LSST than in precursor surveys such as ZTF. The overlap in redshift constitutes a potential difficulty when it comes to distinguishing lensed SNe from unlensed ones in LSST. Nevertheless, Fig.~\ref{fig:colour-mag} demonstrates that we can achieve a better separation by combining colours with apparent magnitudes, since lensed SNe (especially the quads) are magnified and hence brighter than unlensed ones at the same redshifts. We also see a few very red lensed SNe at high redshifts where unlensed SNe are not visible anymore with Rubin (redshifts $\gtrapprox 1$).

\medskip
We investigate each colour and magnitude combination at multiple epochs and obtain the following linear cuts using the method described in Sec.~\ref{subsect:methods:colour-mag}:

\begin{eqnarray*}
g-r &>& 0.84 m_ r -16.9  \\ 
g-i &>& 0.69 m_ i -13.5  \\ 
g-z &>& 1.41 m_ z -29.4  \\ 
g-y &>& 1.14 m_ y -23.6  \\ 
r-i &>& 0.44 m_ i -9.3  \\ 
r-z &>& 0.88 m_ z -19.1  \\ 
r-y &>& 1.0 m_ y -21.7  \\ 
i-z &>& 0.58 m_ z -12.6  \\ 
i-y &>& 0.82 m_ y -17.9  \\ 
z-y &>& 0.15 m_ y -2.9 \, .
\end{eqnarray*}

\begin{figure*}
	\centering
		{\includegraphics[width=\textwidth,clip=true]{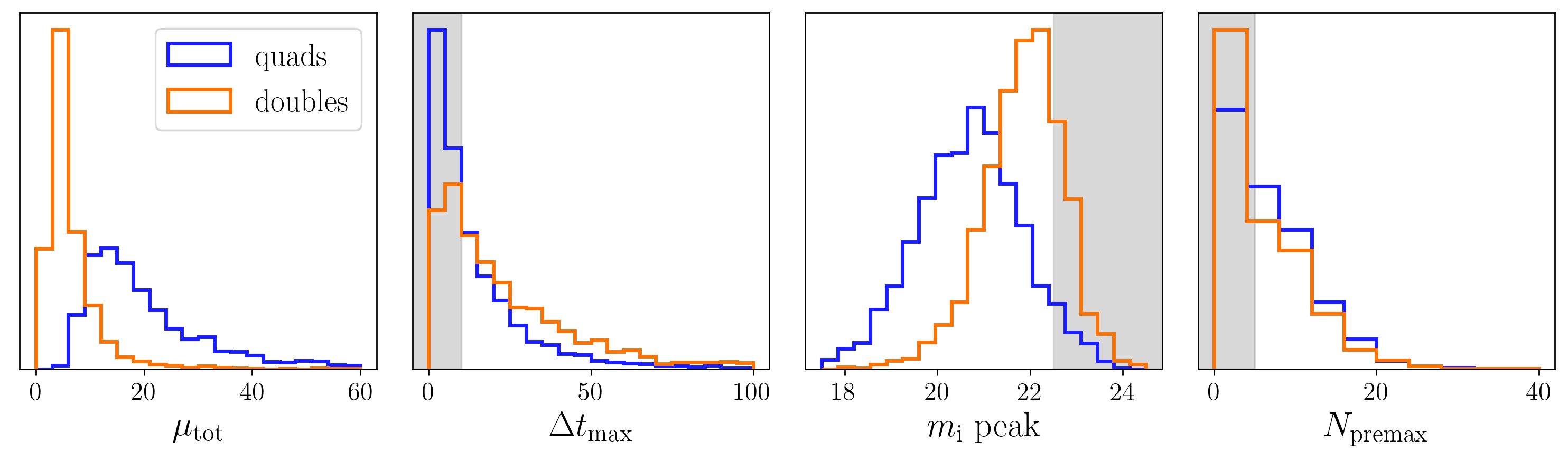}}
	\caption{Normalised distributions showing the properties of doubles and quads from the detected lensed SNIa sample. From left to right, the panels show the total magnification ($\mu_{\rm tot}$), maximum time delay between the images ($\Delta t_{\rm max}$), apparent $i$-band magnitude at peak ($m_{\rm i}$), and the number of detections before peak ($N_{\rm premax}$). The non-shaded regions indicate the objects that satisfy the conditions for the `gold' sample.}
\label{fig:goldsample}
\end{figure*}

\medskip
We calculate the percentage of the lensed SNIa and the percentage of background contaminants from the simulated unlensed population that pass the colour-magnitude cuts. We find that 41\% (43\%) of the lensed doubles (quads) pass at least one of the mentioned colour magnitude cuts. 2-3\% of the unlensed SNIa would also pass this colour-magnitude cut. 
We find that the parameters combination that best separates lensed from unlensed SN for LSST is $r-z$ peak colour versus observed apparent $z$-band peak magnitude, which keeps 23\% (26\%) of the doubles (quads) with only 0.7\% of the unlensed sample. 28\% (32\%) of the lensed doubles (quads) would pass  $r-i$ peak colour versus observed apparent $i$-band peak magnitude, but with a almost 2\% of the unlensed events. \newrevision{Since unlensed SNe outnumber lensed ones by a factor of $\sim 10^3$ we want to keep the false-positive rate low.}
The combination with the lowest contaminants is $r-y$ peak colour versus observed apparent $y$-band peak magnitude, with only 0.1\% unlensed events, but due to the poorer cadence and depth for the $y$-filter, requiring detections in $y$ means we only preserve 15\% (20\%) of doubles (quads). 
As detailed in Table~\ref{tab:results}, this corresponds to around 20 lensed SNIa a year that pass one of the colour-magnitude cuts. 
\mnrasrevision{While colour and magnitude cuts alone are not enough to distinguish completely between lensed and unlensed SNIa, this analysis shows that they can be useful additional tools to inform us about lensed SNIa candidates in LSST.}

\subsection{Time-delay measurements from LSST-only data}
\label{subsect:results:dt_LSST}

For a small fraction of the simulated lensed SNIa, we find that we can extract a useful time-delay measurement using only LSST data. An example of such an object is shown in Fig.~\ref{fig:glSN_dt_saltfit}. However, for most cases the light curve has a sparse sampling such that a \revision{\texttt{SALT3}} fit with constrained $x_1$ and $c$ is unsuccessful. Less than 2\% of our detected lensed SNIa sample allows for a $\Delta t$ measurement \mnrasrevision{with accuracy better than $5\%$}. \mnrasrevision{Additionally, in \citet{Hayes2023} our lensed SNIa simulations are fitted with \texttt{GausSN}\footnote{\href{https://github.com/erinhay/gaussn}{https://github.com/erinhay/gaussn}}, a Bayesian semi-parametric Gaussian Process time-delay estimation model. \texttt{GausSN} recovers for $\sim 13\%$ of the resolved double systems the time delays within $5\%$ accuracy, corresponding to $6\%$ of the total detected sample. These results} emphasise the importance of follow-up observations to improve the quality of the time-delay measurements, \revision{in line with the conclusions from \citet{Huber2019LSSTcadence}}. The number of lensed SNIa a year that qualifies for accurate $\Delta t$ measurements with follow-up observations is discussed in Sec.~\ref{subsect:results:goldsample}.

\subsection{Gold sample and cosmological prospects}
\label{subsect:results:goldsample}

Fig.~\ref{fig:goldsample} shows the early detections, peak magnitude, and time delay distributions of the detected sample. When applying the cuts described in Sec.~\ref{subsect:methods:goldsample}, we find that 25\% of the detected lensed SNIa belong to the `gold' sample. This corresponds to roughly 10 systems per year, as outlined in Table~\ref{tab:results}, for which high-quality follow-up observations and precision cosmology measurements are expected to be feasible. 

\medskip
To estimate the cosmological prospects of such a gold sample, we assume the availability, for each system in the sample, of ground-based follow-up observations to sample the light curve well, and high-resolution imaging and spatially-resolved spectroscopy to constrain both the time delays and the lens mass model. We expect that the uncertainty in the inferred time delay, with high-resolution follow-up, will be $\sim 0.1$--$0.5$ days, as inferred from local SN~Ia samples \citep[e.g.,][]{Johansson2021} corresponding to a conservative upper limit on the time-delay error of 5\%. \revision{ For the lens mass model, we assume an uncertainty of $7\%$.} Since SNe are explosive transients, we can obtain post-explosion images to cross-check the lens model, and hence, reduce the uncertainty in the final lens mass model estimate. 
\mnrasrevision{\citet{ding2021improved} predicts a factor~4 improvement in the lens model for lensed SNe over lensed quasars, although the difference will be smaller for observations at higher angular resolution than the \textit{Hubble Space Telescope}.}
Observations with Integral Field Units (IFUs) can measure the stellar kinematics and help to further break the mass sheet degeneracy \revision{\citep{Birrer&Treu2021}}. \revision{ We note, furthermore, that systems which do not have very precisely measured time delays \citep[e.g., iPTF16geu, SN~Zwicky][]{Dhawan2020_16geu,Goobar2023_SNZwicky} can still be important for reducing uncertainties on the mass modelling, via a precisely measured model independent estimate of the lensing magnification \citep{birrer2021standardizable}}.
Combining the uncertainties from the time-delay measurement and the mass modelling, we \revision{obtain} a precision of \revision{8.6\%} in $H_0$ for each system in the `gold' sample. Consequently, we would need \revision{30} lensed SNIa to reduce the uncertainty to 1.5\% in $H_0$, corresponding to \revision{$\sim 3$} years of LSST observations. Furthermore, we note that lensed SNe with shorter time delays (e.g. $5 < \Delta t < 10$ days) will also contribute to improving the precision, even though the individual uncertainties per system would be greater than \revision{8.6\%}. In that case, we could reach the expected precision in a shorter duration of the LSST survey.

\section{Discussion and conclusions}
\label{sect:conclusions}

\medskip
In this work, we studied the detectability of lensed SNIa in LSST.
We have investigated the impact of the LSST baseline v3.0 observing strategy and of microlensing on the predicted annual lensed SNIa detections. \mnrasrevision{Similar to \citet{wojtak2019magnified}, we consider both the magnification method as used in \citet{goldstein2017glSNe} and image multiplicity method as used in \citet{oguri2010gravitationally} for detecting lensed SNIa. Compared to the aforementioned simulations, we go from limiting magnitude cuts to fully simulated LSST observations including weather effects and stellar microlensing.}
The LSST observing strategy is expected to proceed using a rolling cadence, in which certain areas of the WFD footprint will be assigned more frequent visits than others.
The expected yearly number of lensed SNIa is higher for a non-rolling cadence (50 events) than for a rolling cadence (44 events), but the difference does not appear to be detrimental to the lensed SNe science case. Microlensing effects from stars in the lens galaxy result in a handful fewer detected lensed SNe per year. We found that $\sim 40\%$ of lensed SNIa detected in LSST will stand out from unlensed SNIa with simple linear cuts in colour and peak magnitude. Using only LSST data, a time delay within 5\% of the truth value is expected to be measured for only a small fraction, $\sim 2\%$ of the systems. Hence it is important to assess the feasibility of time-delay and $H_0$ measurements from follow-up observations.

\medskip
We have determined a set of detectability criteria that will allow for timely follow-up and cosmological inference. Our results predict $\sim$ 10 lensed SNIa per year that will have sufficient early detections and will be sufficiently bright for follow-up observations, while also having time delays larger than 10 days to enable time delay cosmography. Assuming uncertainties of \revision{$8.6\%$} in $H_0$ per object, this sample is expected to enable a Hubble constant measurement of $1.5\%$ precision in \revision{three} years of LSST observations.

\medskip
Our results only focus on SNIa; the expected number of lensed core-collapse SNe is likely even higher \citep{wojtak2019magnified,goldstein2017glSNe, oguri2010gravitationally}.
Future work is needed to investigate the cosmological prospects of  strongly-lensed core-collapse SNe, which display more intrinsic variation than type Ias \revision{but could be used efficiently for spectroscopic time-delay measurements (see e.g. \cite{Bayer2021} for type IIP SNe)}. Additionally, in this work we have not investigated other sources of background contamination than unlensed SNIa. \newrevision{Other future studies include testing the resolvable separation by injecting lensed SNe in the LSST difference imaging analysis (DIA) pipeline (e.g. Liu et al., in prep.) and testing the pixel-to-cosmology performance for lensed SNe in an analysis similar to \citet{Sanchez2022}.}

\medskip
We have shown that lensed SNIa discovered with LSST will be excellent precision probes of cosmology. Crucial to the success of this programme is the availability and coordination of follow-up resources, both monitoring of light curves to measure time delays, and high-resolution imaging data and spatially resolved kinematics to constrain the lens mass model. With the selection cuts applied, our work shows that the sample will be sufficiently bright to \revision{measure the present-day expansion rate of the Universe with high precision.}

\section*{Acknowledgements}

This paper has undergone internal review in the LSST
Dark Energy Science Collaboration. \revision{The authors would like to thank Simon Birrer, Philippe Gris, and Peter Nugent for their helpful comments and reviews.}
We are also grateful to Henk Arendse for database advice, 
Christian Setzer for help with simulations,
and Edvard Mörtsell, D'Arcy Kenworthy, Richard Kessler, \mnrasrevision{and Luke Weisenbach} for useful discussions.

\medskip

\noindent Author contributions are listed below. \\
NA: conceptualization, methodology, software (lensed SN simulations), formal analysis, writing (original draft; review \& editing), visualization; \\
SD: methodology, software, validation, formal analysis (time-delay measurements and gold sample), writing (original draft), visualization; \\
ASC: methodology, software, validation, formal analysis (colour-magnitude diagrams), writing (original draft), visualization; \\
HVP: conceptualization, validation, writing (review \& editing), supervision, funding acquisition; \\
AG: validation, writing (review), funding acquisition; \\
RW: software, validation, writing (review); \\
CA: validation; \\
RB: software (OpSim Summary), writing (original draft); \\
SH: software (microlensing simulations), writing (original draft); \\
\newrevision{SB: writing (collaboration internal review)}

\medskip
This work has been enabled by support from the research project grant ‘Understanding the Dynamic Universe’ funded by the Knut and Alice Wallenberg Foundation under Dnr KAW 2018.0067. This project has received funding from the European Research Council (ERC) under the European Union’s Horizon 2020 research and innovation programme (grant agreement no. 101018897 CosmicExplorer). The work of HVP was partially supported by the G{\"o}ran Gustafsson Foundation for Research in Natural Sciences and Medicine. This work was partially enabled by funding from the UCL Cosmoparticle Initiative. SD acknowledges support from the Marie Curie Individual Fellowship under grant ID 890695 and a Junior Research Fellowship at Lucy Cavendish College. \newrevision{SB thanks Stony Brook University for support.}

\medskip
The DESC acknowledges ongoing support from the
Institut National de Physique Nucléaire et de Physique
des Particules in France; the Science \& Technology Facilities Council in the United Kingdom; and the Department of Energy, the National Science Foundation, and
the LSST Corporation in the United States. DESC uses
resources of the IN2P3 Computing Center (CC-IN2P3–
Lyon/Villeurbanne - France) funded by the Centre National de la Recherche Scientifique; the National Energy
Research Scientific Computing Center, a DOE Office
of Science User Facility supported by the Office of Science of the U.S. Department of Energy under Contract
No. DE-AC02-05CH11231; STFC DiRAC HPC Facilities, funded by UK BEIS National E-infrastructure capital grants; and the UK particle physics grid, supported by the GridPP Collaboration. This work was performed in part under DOE Contract DE-AC02-76SF00515.

\medskip
\revision{
\textit{Software:} Astropy \citep{Astropy2013, Astropy2018, Astropy2022},
Jupyter \citep{JupyterNotebook},
Matplotlib \citep{Matplotlib2007, Matplotlib2020},
NumPy \citep{Numpy202},
Pandas \citep{Pandas2010, Pandas2023},
Pickle \citep{Pickle2020},
SciPy \citep{Scipy2020},
Seaborn \citep{Seaborn2020},
SQLite \citep{sqlite2020hipp},
ChainConsumer \citep{Hinton2016},
Lenstronomy \citep{birrer2018lenstronomy, Birrer2021LenstronomyII},
SNcosmo \citep{barbary2016sncosmo},
SALT3 \citep{guy2007salt2, Kenworthy2021},
OpSimSummary \citep{2020ApJS..247...60B}.}

\section*{Data Availability}

The simulation code \textit{lensed Supernova Simulator Tool} (\texttt{lensedSST}) and lensed SNIa catalogues are publicly available at \url{https://github.com/Nikki1510/lensed_supernova_simulator_tool}.



\bibliographystyle{mnras}
\bibliography{./compiled_references}



\appendix
\onecolumn

\section{Colour and magnitude cuts}
\label{appendix:colour-mag}

\begin{figure}
    \centering
    \includegraphics[width=\textwidth]{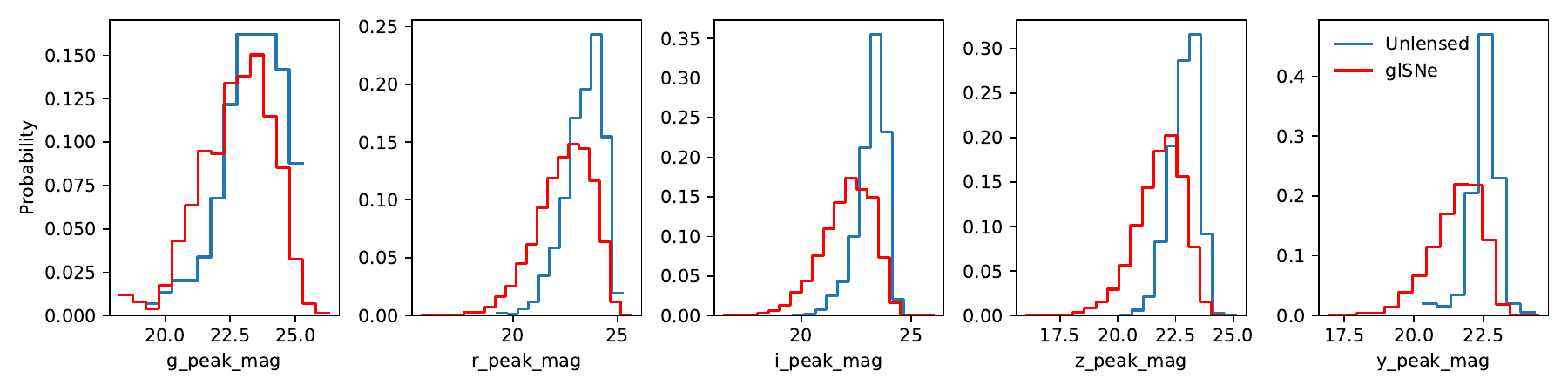}
    \caption{Peak apparent magnitudes in the $g, r, i, z, y$-bands for simulated lensed and unlensed SNIa.}
    \label{appendixfig:mag_all}
\end{figure}

\begin{figure}
    \centering
    \includegraphics[width=\textwidth]{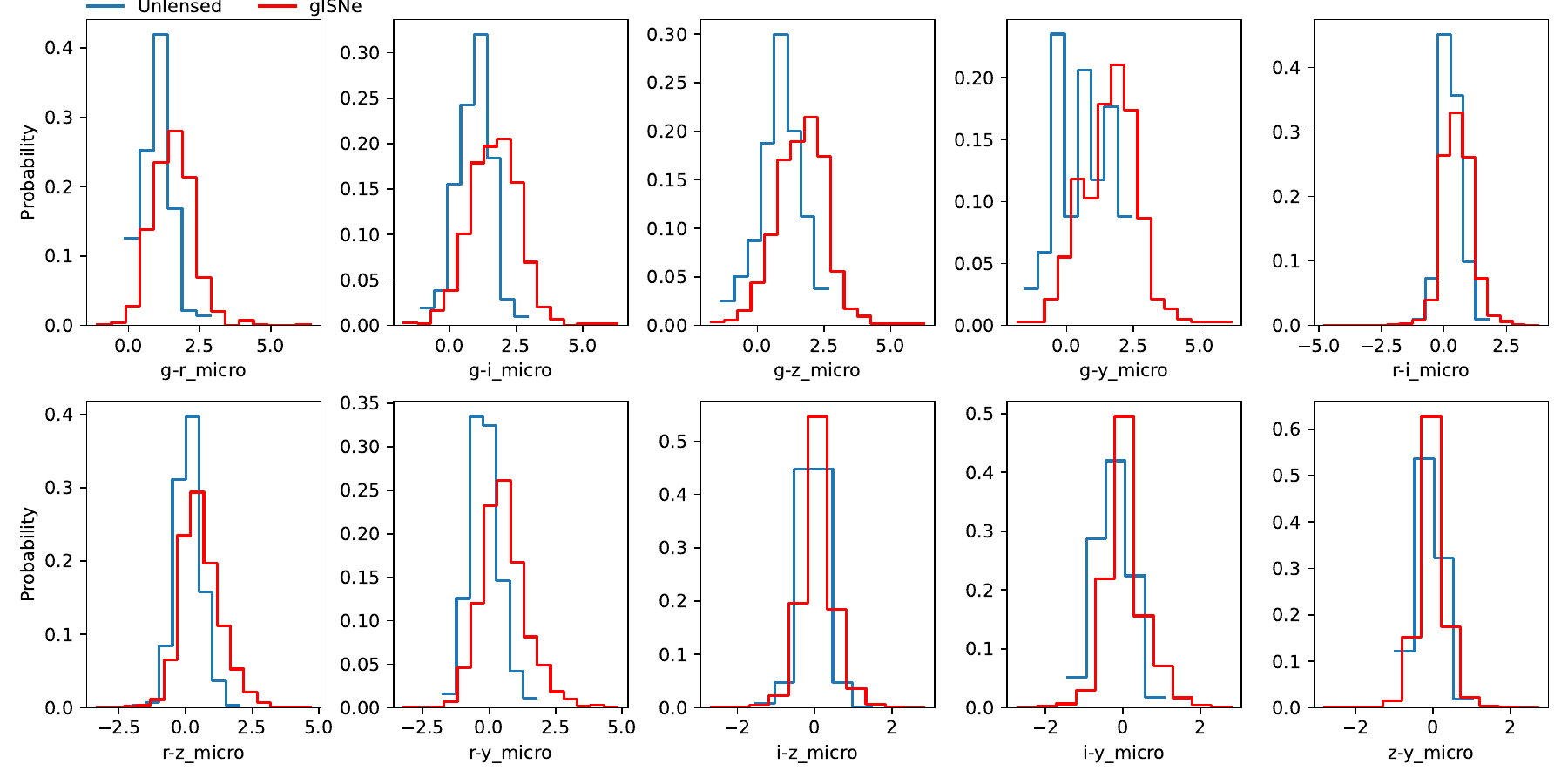}
    \caption{Colours ($g-r$, $g-i$, $g-z$, $g-y$,$r-i$, $r-z$, $r-y$, $i-z$, $i-y$ and $z-y$) for simulated lensed and unlensed SNIa.}
    \label{appendixfig:colour_all}
\end{figure}

\begin{figure}
    \centering
    \includegraphics[width=\textwidth]{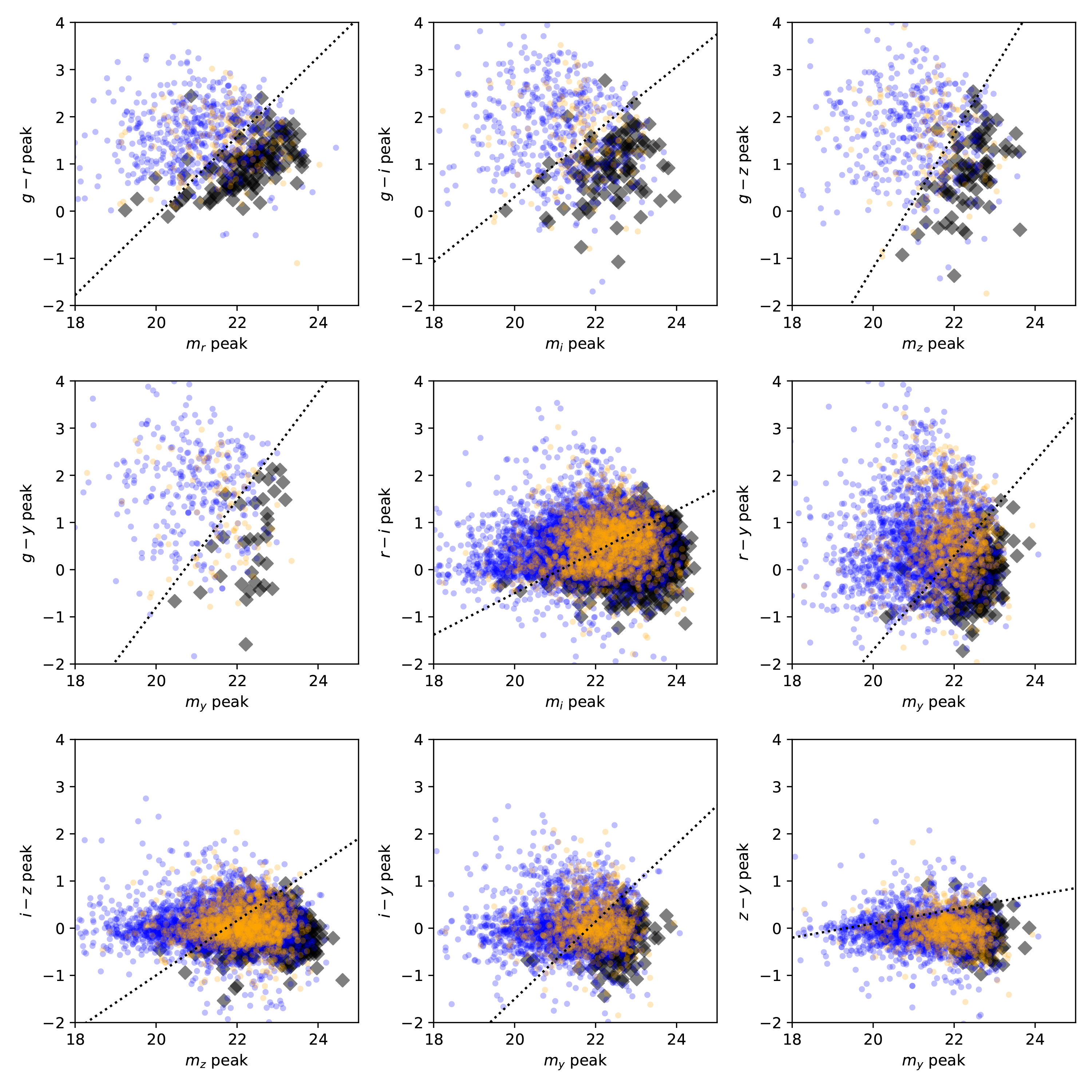}
    \caption{Colour-magnitude cuts described in Sec.~\ref{subsect:results:colour_mag} used (together with $r-z$ versus $z$ shown in Fig.~\ref{fig:colour-mag}) to calculate the total number of lensed events that are distinguishable from unlensed SNIa.}
    \label{appendixfig:colour-mag}
\end{figure}

Here, we provide all results from the colour-magnitude investigation of simulated lensed and unlensed SNIa as described in Sec.~\ref{subsect:methods:colour-mag}. Fig.~\ref{appendixfig:mag_all} shows the 1D distributions of peak apparent magnitudes in the $g,r,i,z$-bands and Fig.~\ref{appendixfig:colour_all} the colours at peak. The joint colour-magnitude diagrams are displayed in Fig.~\ref{appendixfig:colour-mag}. Our results predict that the colour-magnitude combination with the strongest separation between lensed and unlensed SNIa is $r-z$ colours versus $z$-band magnitudes, but we combine all colour-magnitude combinations for the best result.

\bsp
\label{lastpage}
\end{document}